\documentclass[letterpaper,twocolumn,10pt]{article}
\usepackage{usenix-2020-09}

\usepackage{color}
\usepackage{tabularray}
\UseTblrLibrary{siunitx}
\usepackage{array}

\usepackage{pifont}
\usepackage{xspace}
\usepackage{amssymb}
\usepackage{amsmath,amsthm}
\usepackage{booktabs,tabularx}
\usepackage{amsfonts}
\usepackage{enumitem}
\usepackage{xcolor,colortbl}
\usepackage[small]{caption}
\usepackage{subfig}
\usepackage{fp}

\usepackage[skins]{tcolorbox}

\usepackage{textcomp}

\usepackage{wasysym}
\usepackage{nicematrix}
\usepackage{fontawesome5}

\usepackage{listings}
\definecolor{mygreen}{rgb}{0,0.6,0}
\definecolor{mymauve}{rgb}{0.58,0,0.82}
\definecolor{mygray}{gray}{0.95}
\definecolor{mygray1}{gray}{0.6}

\usepackage{subfig}
\usepackage{xcolor}
\usepackage{graphicx}
\usepackage{pgf-umlsd}
\usepackage{tikz}
\usepackage{cite}
\usepackage{standalone}
\usepackage{multirow}
\usepackage{adjustbox}
\usepackage{wrapfig}

\usetikzlibrary{positioning}
\usetikzlibrary{shapes,arrows,calc,automata}
\usetikzlibrary{shapes.geometric}

\usepackage{hyperref}
\definecolor{BrickRed}{HTML}{B6321C}
\hypersetup{
  colorlinks   = true,
  urlcolor     = blue, 
  linkcolor    = blue, 
  citecolor   = blue
}

\usepackage{url}

\usepackage{breakurl}
\usepackage{booktabs}

\usepackage[nameinlink]{cleveref}
\Crefname{requirement}{Req.}{Reqs.}
\Crefname{equation}{Eq.}{Eqs.}
\Crefname{figure}{Fig.}{Figs.}
\Crefname{tabular}{Tab.}{Tabs.}
\Crefname{section}{Sec.}{Sec
encircle.}

\usepackage{multirow}

\usepackage{cleveref}

\definecolor{yl}{HTML}{FCF803}

\newcommand*\circled[1]{\tikz[baseline=(char.base)]{
            \node[shape=circle,draw,inner sep=0.3pt,font=\bf\small] (char) {#1};}}
            
\newcommand*\circledy[1]{\tikz[baseline=(char.base)]{
            \node[shape=circle,draw,inner sep=0.3pt,fill=yl,font=\bf\small] (char) {#1};}}

\newif \ifPPI
\PPIfalse

\newif \ifARXIV
\ARXIVtrue

\newif \ifSendETH
\SendETHfalse

\usepackage{graphicx}
\usepackage{tcolorbox}
\usepackage{array}
\newcolumntype{N}{@{}m{0pt}@{}}

\newcounter{observationCounter}
\crefname{observationCounter}{observation}{observations}
\Crefname{observationCounter}{Observation}{Observations}

\newtcolorbox[auto counter]{observation-box}[2][]
{%
  left skip = 0cm,
  size = small,%
  before upper=\par\noindent{},
  colframe = black,%
  colback  = blue!5!white,%
  coltitle = white,%
  title    = {#2},%
  #1,%
  enhanced,%
}

\newtcolorbox[auto counter]{observation-box-new}[2][]
{%
attach title to upper,after title={:\ },
size = small,%
left skip = 0cm,
colbacktitle=red!10!white,
colback= blue!5!white,
coltitle=black,
title={#2},
fonttitle=\bfseries,#1
}

\newcounter{requirementCounter}
\newenvironment{requirement}[0]
  {\par\addvspace{1\baselineskip}%
   \refstepcounter{requirementCounter}%
   {\noindent$\rightarrow$ \emph{Requirement \therequirementCounter}: }}

\crefname{requirementCounter}{requirement}{requirements}
\Crefname{requirementCounter}{Requirement}{Requirements}

\crefname{invariantCounter}{invariant}{invariant}
\Crefname{invariantCounter}{Invariant}{Invariant}

\newcounter{myctr}

\newenvironment{mylist}
    {\begin{list}{(\textbf{\arabic{myctr}})}
        {\usecounter{myctr}
        \setlength{\topsep}{0mm}\setlength{\itemsep}{0.5mm}
        \setlength{\parsep}{0.5mm}
        \setlength{\itemindent}{0mm}\setlength{\partopsep}{0mm}
        \setlength{\labelwidth}{-2mm}
        \setlength{\leftmargin}{1mm}}
    }
    {\end{list}}

\newenvironment{mybullet}
    {\begin{list}{$\bullet$}
        {\setlength{\topsep}{0mm}\setlength{\itemsep}{0.5mm}
        \setlength{\parsep}{0.5mm}
        \setlength{\itemindent}{0mm}\setlength{\partopsep}{0mm}
        \setlength{\labelwidth}{-2mm}
        \setlength{\leftmargin}{1mm}}
    }
    {\end{list}}

\newcommand{\myparagraph}[1]{\noindent\textbf{#1.}}

\newcommand{\spacesave}{\vspace{-10pt}}

\let\oldding\ding
\renewcommand{\ding}[2][1]{\scalebox{#1}{\oldding{#2}}}

\newcommand{\one}{\ding[1.2]{172}\xspace}
\newcommand{\two}{\ding[1.2]{173}\xspace}
\newcommand{\three}{\ding[1.2]{174}\xspace}
\newcommand{\four}{\ding[1.2]{175}\xspace}
\newcommand{\five}{\ding[1.2]{176}\xspace}
\newcommand{\six}{\ding[1.2]{177}\xspace}
\newcommand{\seven}{\ding[1.2]{178}\xspace}
\newcommand{\eight}{\ding[1.2]{179}\xspace}
\newcommand{\nine}{\ding[1.2]{180}\xspace}
\newcommand{\ten}{\ding[1.2]{181}\xspace}

\newcommand{\req}{\CIRCLE}
\newcommand{\partreq}{\RIGHTcircle}
\newcommand{\notreq}{\Circle}

\newcommand{\supported}{\ding{51}}
\newcommand{\notsupported}{\ding{55}}

\definecolor{mygreen1}{RGB}{169, 209, 142}
\definecolor{myyellow1}{RGB}{255, 230, 153}
\definecolor{myblue1}{RGB}{180, 199, 231}

\newcommand{\splitatcommas}[1]{%
  \begingroup
  \begingroup\lccode`~=`, \lowercase{\endgroup
    \edef~{\mathchar\the\mathcode`, \penalty0 \noexpand\hspace{0pt plus 1em}}%
  }\mathcode`,="8000 #1%
  \endgroup
}

\newcolumntype{?}{!{\vrule width 1pt}}

\theoremstyle{definition}

\newcommand{\ascendcc}{\textsc{Ascend-CC}}

\newcolumntype{R}[2]{%
    >{\adjustbox{angle=#1,lap=\width-(#2)}\bgroup}%
    l%
    <{\egroup}%
}

\definecolor{Gray}{gray}{0.85}

\newcolumntype{a}{>{\columncolor{Gray}}c}
\newcolumntype{b}{>{\columncolor{white}}c}

\usepackage[
all=normal,floats=tight
,paragraphs=tight
,wordspacing=tight
,mathspacing=tight
,mathdisplays=tight
,indent=tight
]{savetrees}

\colorlet{punct}{red!60!black}
\definecolor{background}{gray}{0.99}
\definecolor{delim}{RGB}{20,105,176}
\colorlet{numb}{magenta!60!black}
\definecolor{eclipseStrings}{RGB}{42,0.0,255}
\definecolor{eclipseKeywords}{RGB}{127,0,85}
\definecolor{codegreen}{rgb}{0,0.6,0}

\lstdefinelanguage{json}{
    basicstyle=\bfseries\scriptsize\ttfamily,
    showstringspaces=false,
    breaklines=true,
    frame=lines,
    backgroundcolor=\color{background},
    morekeywords={TRUE,FALSE,linkEnc},
    keywordstyle=\color{numb},
    literate=
     *{0}{{{\color{numb}0}}}{1}
      {1}{{{\color{numb}1}}}{1}
      {2}{{{\color{numb}2}}}{1}
      {3}{{{\color{numb}3}}}{1}
      {4}{{{\color{numb}4}}}{1}
      {5}{{{\color{numb}5}}}{1}
      {6}{{{\color{numb}6}}}{1}
      {7}{{{\color{numb}7}}}{1}
      {8}{{{\color{numb}8}}}{1}
      {9}{{{\color{numb}9}}}{1}
      {:}{{{\color{punct}{:}}}}{1}
      {,}{{{\color{punct}{,}}}}{1}
      {\{}{{{\color{delim}{\{}}}}{1}
      {\}}{{{\color{delim}{\}}}}}{1}
      {[}{{{\color{delim}{[}}}}{1}
      {]}{{{\color{delim}{]}}}}{1},
}

\lstdefinelanguage{ccode}{
    basicstyle=\bfseries\footnotesize\ttfamily,
    showstringspaces=false,
    numbers=left,
    commentstyle=\color{codegreen},
    language=C,
    breaklines=true,
    frame=lines,
    morecomment=[f][\color{green}][0]{*},
    morecomment=[f][\color{red}][0]{\#},
    backgroundcolor=\color{background},
    morekeywords={TRUE,false,main, execute,memcpy,context,loadmodel},
    keywordstyle=\color{numb},
    literate=
     *{0}{{{\color{numb}0}}}{1}
      {1}{{{\color{numb}1}}}{1}
      {2}{{{\color{numb}2}}}{1}
      {3}{{{\color{numb}3}}}{1}
      {4}{{{\color{numb}4}}}{1}
      {5}{{{\color{numb}5}}}{1}
      {6}{{{\color{numb}6}}}{1}
      {7}{{{\color{numb}7}}}{1}
      {8}{{{\color{numb}8}}}{1}
      {9}{{{\color{numb}9}}}{1}
      {:}{{{\color{punct}{:}}}}{1}
      {,}{{{\color{punct}{,}}}}{1}
      {\{}{{{\color{delim}{\{}}}}{1}
      {\}}{{{\color{delim}{\}}}}}{1}
      {[}{{{\color{delim}{[}}}}{1}
      {]}{{{\color{delim}{]}}}}{1},
}

\DeclareFixedFont{\ttb}{T1}{txtt}{bx}{n}{9} 
\DeclareFixedFont{\ttm}{T1}{txtt}{m}{n}{9}  
\definecolor{deepblue}{rgb}{0,0,0.5}
\definecolor{deepred}{rgb}{0.6,0,0}
\definecolor{deepgreen}{rgb}{0,0.5,0}

\newcommand\pythonstyle{\lstset{
language=Python,
basicstyle=\ttm,
morekeywords={self},              
keywordstyle=\ttb\color{deepblue},
emph={MyClass,__init__},          
emphstyle=\ttb\color{deepred},    
stringstyle=\color{deepgreen},
frame=tb,                         
showstringspaces=false
}}

\lstnewenvironment{python}[1][]
{
\pythonstyle
\lstset{#1}
}
{}

\newcommand\pythoninline[1]{{\pythonstyle\lstinline!#1!}}

\definecolor{maroon}{cmyk}{0, 0.87, 0.68, 0.32}
\definecolor{halfgray}{gray}{0.55}
\definecolor{ipython_frame}{RGB}{207, 207, 207}
\definecolor{ipython_bg}{RGB}{247, 247, 247}
\definecolor{ipython_red}{RGB}{186, 33, 33}
\definecolor{ipython_green}{RGB}{0, 128, 0}
\definecolor{ipython_cyan}{RGB}{64, 128, 128}
\definecolor{ipython_purple}{RGB}{170, 34, 255}

\usepackage{listings}
\lstdefinelanguage{iPython}{
    morekeywords={access,and,break,class,continue,def,del,elif,else,except,exec,finally,for,from,global,if,import,in,is,lambda,not,or,pass,print,raise,return,try,while},%
    morekeywords=[2]{abs,all,any,basestring,bin,bool,bytearray,callable,chr,classmethod,cmp,compile,complex,delattr,dict,dir,divmod,enumerate,eval,execfile,file,filter,float,format,frozenset,getattr,globals,hasattr,hash,help,hex,id,input,int,isinstance,issubclass,iter,len,list,locals,long,map,max,memoryview,min,next,object,oct,open,ord,pow,property,range,raw_input,reduce,reload,repr,reversed,round,set,setattr,slice,sorted,staticmethod,str,sum,super,tuple,type,unichr,unicode,vars,xrange,zip,apply,buffer,coerce,intern},%
    sensitive=true,%
    morecomment=[l]\#,%
    morestring=[b]',%
    morestring=[b]",%
    morestring=[s]{'''}{'''},
    morestring=[s]{"""}{"""},
    morestring=[s]{r'}{'},
    morestring=[s]{r"}{"},%
    morestring=[s]{r'''}{'''},%
    morestring=[s]{r"""}{"""},%
    morestring=[s]{u'}{'},
    morestring=[s]{u"}{"},%
    morestring=[s]{u'''}{'''},%
    morestring=[s]{u"""}{"""},%
    literate=
    {á}{{\'a}}1 {é}{{\'e}}1 {í}{{\'i}}1 {ó}{{\'o}}1 {ú}{{\'u}}1
    {Á}{{\'A}}1 {É}{{\'E}}1 {Í}{{\'I}}1 {Ó}{{\'O}}1 {Ú}{{\'U}}1
    {à}{{\`a}}1 {è}{{\`e}}1 {ì}{{\`i}}1 {ò}{{\`o}}1 {ù}{{\`u}}1
    {À}{{\`A}}1 {È}{{\'E}}1 {Ì}{{\`I}}1 {Ò}{{\`O}}1 {Ù}{{\`U}}1
    {ä}{{\"a}}1 {ë}{{\"e}}1 {ï}{{\"i}}1 {ö}{{\"o}}1 {ü}{{\"u}}1
    {Ä}{{\"A}}1 {Ë}{{\"E}}1 {Ï}{{\"I}}1 {Ö}{{\"O}}1 {Ü}{{\"U}}1
    {â}{{\^a}}1 {ê}{{\^e}}1 {î}{{\^i}}1 {ô}{{\^o}}1 {û}{{\^u}}1
    {Â}{{\^A}}1 {Ê}{{\^E}}1 {Î}{{\^I}}1 {Ô}{{\^O}}1 {Û}{{\^U}}1
    {œ}{{\oe}}1 {Œ}{{\OE}}1 {æ}{{\ae}}1 {Æ}{{\AE}}1 {ß}{{\ss}}1
    {ç}{{\c c}}1 {Ç}{{\c C}}1 {ø}{{\o}}1 {å}{{\r a}}1 {Å}{{\r A}}1
    {€}{{\EUR}}1 {£}{{\pounds}}1
    {^}{{{\color{ipython_purple}\^{}}}}1
    {=}{{{\color{ipython_purple}=}}}1
    {+}{{{\color{ipython_purple}+}}}1
    {*}{{{\color{ipython_purple}$^\ast$}}}1
    {/}{{{\color{ipython_purple}/}}}1
    {+=}{{{+=}}}1
    {-=}{{{-=}}}1
    {*=}{{{$^\ast$=}}}1
    {/=}{{{/=}}}1,
    literate=
    *{-}{{{\color{ipython_purple}-}}}1
     {?}{{{\color{ipython_purple}?}}}1,
    identifierstyle=\color{black}\ttfamily,
    commentstyle=\color{ipython_cyan}\ttfamily,
    stringstyle=\color{ipython_red}\ttfamily,
    keepspaces=true,
    showspaces=false,
    showstringspaces=false,
    rulecolor=\color{ipython_frame},
    frame=single,
    frameround={t}{t}{t}{t},
    framexleftmargin=6mm,
    numbers=left,
    numberstyle=\tiny\color{halfgray},
    backgroundcolor=\color{ipython_bg},
    basicstyle=\scriptsize,
    keywordstyle=\color{ipython_green}\ttfamily,
}

\begin{document}

\date{}

\title{\Large \bf \ascendcc: Confidential Computing on Heterogeneous NPU \\ for Emerging Generative AI Workloads}

\ifARXIV
\author{\rm{Aritra Dhar$^{\bigstar\dag}$ \qquad Clément Thorens$^{\ddag\dag}$\thanks{Work done while the author was in Huawei Zurich Research Center} \qquad Lara Magdalena Lazier$^{\bigstar}$ \qquad Lukas Cavigelli$^{\bigstar}$}  \\ \\
$^\dag$ Joint first authors
\\ \\
$^\bigstar${Huawei Zurich Research Center} ~\;~\;~\;~\;~\;~\;~\;~\;~\;~\; $^{\ddag}${ETH Zurich} \\ \\
}

\date{}
\fi

\maketitle

\begin{abstract}

Cloud workloads have dominated generative AI based on large language models (LLM). 
Specialized hardware accelerators, such as GPUs, NPUs, and TPUs, play a key role in AI adoption due to their superior performance over general-purpose CPUs. 
The AI models and the data are often highly sensitive and come from mutually distrusting parties.
Existing CPU-based TEEs such as Intel SGX or AMD SEV do not provide sufficient protection.
Device-centric TEEs like Nvidia-CC only address tightly coupled CPU-GPU systems with a proprietary solution requiring TEE on the host CPU side. 
On the other hand, existing academic proposals are tailored toward specific CPU-TEE platforms.

To address this gap, we propose \ascendcc{}, a confidential computing architecture based on discrete NPU devices that requires no trust in the host system.
\ascendcc{} provides strong security by ensuring data and model encryption that protects not only the data but also the model parameters and operator binaries.
\ascendcc{} uses delegation-based memory semantics to ensure isolation from the host software stack, and task attestation provides strong model integrity guarantees. 
Our \ascendcc{} implementation and evaluation with state-of-the-art LLMs such as Llama2 and Llama3 shows that \ascendcc{} introduces minimal overhead with no changes in the AI software stack.


\end{abstract}

\section{Introduction}
\label{sec:introduction}

Recently, Generative AI (GenAI) has gained momentum, with large language models (LLMs) being used in applications, such as chat bot~\cite{chatgpt}, image and video generation~\cite{dalle,sora}, code completion~\cite{visualstudioGitHubCopilot}.
The GenAI is considered a build block~\cite{zhang2023one} for future artificial general intelligence (AGI).
Major cloud providers offer AI-centric services~\cite{googleInfrastructureModel,microsoftAzureOpenAI,huaweicloudAscendCloud,alibabacloudAlibabaCloud}, that typically utilize specialized accelerators such as GPUs, NPUs, and TPUs.

\myparagraph{Security concerns} These large GenAI workloads bring numerous security challenges in data center environments.
The massive data and computation resources required to train LLMs make them exceedingly expensive~\cite{wiredOpenAIsSays} and prime intellectual properties for the model providers.
Second, the users' queries to the LLMs often contain sensitive information such as health data, personal information, or even business secrets~\cite{forbesSamsungBans}.
In the data center deployments, there are three mutually distrusting parties: the model provider, the data provider, and the cloud provider. 
The \textit{model provider} develops and owns the AI model.
The \textit{data provider} gives the data to the AI model for processing.
Finally, the \textit{cloud provider} owns the computing infrastructure where the AI models are trained or run inference.
The data provider interacts with the computing infrastructure to deliver their data to the AI model.
Therefore, the AI model and data require protection from the cloud provider and the software stack.

\myparagraph{Gap in prior works} Existing CPU-based trusted execution environments (TEE)~\cite{sgx,SEVVM,TZOS,cca} enables secure applications or enclaves isolated from a malicious software stack such as OS/hypervisor.
They can often withstand untrusted DRAM and bus by employing memory encryption.
Outside CPUs, devices such as GPU~\cite{h100,volos2018graviton,jang2019HIX}, IPU~\cite{vaswani2023confidential}, FPGA~\cite{zeitouni2021trusted,zhao2021shef,oh2021meetgo}, provides TEEs. 
Several existing works~\cite{sridhara2023acai,wang2024cage,fen2024trusted,deng2022strongbox,park2023safe} also extend CPU TEE's security primitives to connected devices.
However, except for very few proposals~\cite{vaswani2023confidential}, most existing proposals require a CPU TEE, which increases the Trusted Computing Base (TCB).
Recent side channel attacks~\cite{lipp2018meltdown,foreshadow-usenix18,moghimi2017cachezoom,brasser2017software,paccagnella2021bus} demonstrate that attackers can undermine the security guarantees of CPU TEEs.
Additionally, proposals requiring a confidential VM or C-VM (such as CCA and SEV) further increase the TCB by trusting the C-VM OS. 
Several proposals only consider integrated GPU~\cite{deng2022strongbox,wang2024cage,park2023safe} or NPU~\cite{fen2024trusted} that simplify the problem by not considering PCIe communication and separate memory spaces from CPU.
Placing the part of the driver inside a high-privilege trusted security monitor~\cite{deng2022strongbox,fen2024trusted,10.1145/3492321.3519565} offloads critical security decisions such as memory allocation for tasks and binaries, memory sharing and access control to a trusted entity.
Such designs expand the TCB and do not fit well in a scenario where the host is fully malicious.
While a proposal on GraphCore IPU~\cite{vaswani2023confidential} can withstand a TEE-less host, it does not support a modern AI software stack with interactive sessions with the accelerator and requires extensive hardware modification. 
Finally, almost all the existing TEE proposals consider older CNN-based AI models or evaluate using operators such as matrix multiplication or SVD, which has a small memory footprint.
Therefore, these solutions do not scale to LLMs with large memory footprints and require low latency response.

\myparagraph{Our contribution} 
We design \ascendcc{}, a confidential computing solution for discrete NPUs without relying on a CPU TEE. 
The TCB of \ascendcc{} is only the NPU itself, and the entire host is untrusted.
The hardware root of trust (HW-RoT) in the NPU facilitates key derivation to establish a secure channel between the model and the data provider and enable attestation.
Measured boot ensures the NPU is booted with the correct firmware signed by the hardware manufacturer. 
\ascendcc{} accepts fully encrypted data and models from the data and model provider.
During inference, the NPU removes all the DMA mapping from the host's virtual address space (by removing SMMU entries) to prevent malicious DMA operations from entering its model, data, and workspace (operator execution space).
Only after the memory unmapping does the data and model decryption start.
The results are encrypted before the corresponding memory is DMA mapped to the host.
The NPU runtime creates tasks from the AI models that dictate the order of operator execution (e.g., matrix multiplication or ReLU) and memory operation (such as DMA copy from host to device).
The malicious host can inject tasks (e.g., performing a DMA copy) into the model to compromise the confidentiality of the data and model.
\ascendcc{} performs the task and binary attestation of the model before the model starts executing to ensure that the integrity of the task and model is preserved.
The isolation and end-to-end encryption are provided without introducing changes to AI software stack such as PyTorch.
Therefore, \ascendcc{} does not burden the AI programmer to introduce confidential computing to the existing codebase.

We demonstrate \ascendcc{} on a Huawei Ascend 910A, a state-of-the-art NPU, by modifying its firmware. 
We evaluate \ascendcc{} with state-of-the-art transformer-based LLMs such as GPT-neo-125M, Llama-2-7B-Base, Llama-2-7B-chat, Llama-2-13B-instruct, Llama-3-8B-Base, Llama3-ChatQA-1.5-8B, Llama-3-8B-Instruct, CodeLlama-7B-Instruct, for tasks like chat, sentence completion, and code completion.
Our evaluation shows that \ascendcc{} introduces minimal performance loss during the inference pass(0.91\% in GPT2-neo-125M and 0.028\% in Llama3-Chat-QA-1.5-8B model, both with 2K input sequence size) and one-time set-up. 
Even though our proposal is focused on specific NPU implementation, our design philosophy extends to other AI accelerators, such as GPUs and TPUs that exhibit task-based execution models.

In summary, this paper makes the following contributions:

\begin{mylist} 
    \item \textbf{Identify fundamental building blocks for NPU-based confidential computing.} We identify security properties for protecting models and data from untrusted hosts and cloud providers. 
    Specifically, how to thwart the privileged host from accessing data and models on the NPU and change their memory mapping.
    We list these observations and requirements in~\Cref{sec:motivation_problem_statement}.
    
    \item \textbf{System design and analysis.} We design \ascendcc{} that enables confidential computing in NPU. 

    \item \textbf{End-to-end evaluation.} We implement and evaluate \ascendcc{} with state-of-the-art LLMs and show that \ascendcc{} introduced minimal overhead with no changes to the AI software stack.
\end{mylist}




    

\section{NPU Background}
\label{sec:npu_background}


\begin{figure}[t]
    \centering
    \includegraphics[trim=0cm 9.9cm 16.4cm 0cm, clip,width=0.85\linewidth]{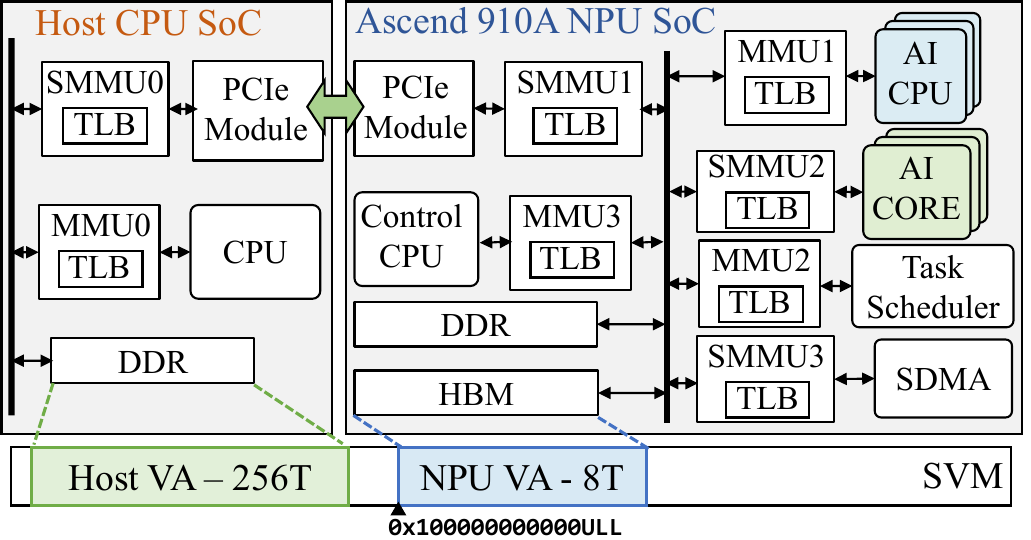}
    \caption{The figure shows a high-level architecture of Ascend 910A SoC along with the shared virtual memory with a 64-bit host CPU. 
    }
    \label{fig:npu_arch}
\end{figure}


\myparagraph{NPU hardware architecture}
In this paper, we focus on the Huawei Atlas 300T card~\cite{huaweiAtlas300T}, which is based on Ascend 910A SoC. 
The Ascend 910A is a state-of-the-art AI SoC designed primarily for large data centers and clouds for training and inference accelerators.
\Cref{fig:npu_arch} shows a high-level view of Ascend 910 NPU SoC architecture. 
All the components of the SoC are connected via an internal bus. 
The NPU SoC has two types of computation cores to execute AI tasks: AI CPU and AI Core.
Four AI CPU cores are general-purpose Huawei Taishan (ARM A73 profile) with hardware cryptographic extensions. 
32 AI Cores are based on the Huawei DaVinci~\cite{liao2021ascend} architecture optimized for executing neural network operations. 
The control CPU is a Taishan core (similar to the AI CPU cores) that runs the NPU firmware and manages the PCIe interfaces. 
After the NPU powers on, the control CPU boots (measured boot) a minimal Linux kernel and initializes the hardware components. 
The task scheduler combines a dedicated hardware component with firmware running on a Taishan core. It distributes AI tasks to the NPU's computing resources, AI CPUs, and AI Cores. 

\myparagraph{NPU driver, runtime, and AI stack}
The NPU runtime stack converts instructions from the higher-level AI software stack (PyTorch/TensorFlow) and communicates with the NPU driver. 
The NPU driver is a set of Linux kernel modules that manages the communications over DMA, issues MMIO commands to send instructions, and monitors NPU health.
Ascend PyTorch adapter (\texttt{torch\_npu}\cite{githubGitHubAscendpytorch}) provides the necessary interfaces to bridge high-level AI-specific APIs to lower-level NPU driver calls.
PyTorch provides two types of model executions: eager and dynamo. 
Eager is an interactive execution in which the tasks are executed as soon as they arrive at the NPU.
This is done with  \texttt{compileAndExecute} instruction that compiles a single operation to a graph and executes it on the NPU.
Dynamo, or graph mode, waits for the entire graph to be compiled and deployed to the NPU before executing the tasks.
The NPU runtime copies the data and model to the NPU HBM. 
The model includes the parameters associated with each AI model layer and corresponding operator binaries (such as matrix multiplication, ReLU, etc.). 
Next, the NPU runtime creates a set of tasks, such as matrix multiplication, to be executed on the AI CPUs or AI Cores.
A task contains operator metadata such as the location of the operator binary on the NPU memory (\texttt{PC\_START}), location of the data arguments, and workspace to store intermediate variables.
The runtime sends the tasks to the task buffer, a reserved NPU memory location.
After sending all tasks, the NPU sends the \texttt{executeModel} task, or, \texttt{compileAndExecute} for Eager mode.
This triggers the task scheduler to read the task buffer in order, select the first task, and submit it to the corresponding AI CPU or AI Core.
After executing the task, the scheduler moves the task entry to the completion queue (CQ) and continues till the task buffer is empty.
The NPU runtime can read the CQ to know the current progress of the execution.
The above description is for a single PCI stream context, and multiple such contexts may exist concurrently.
These streams are processed by the task scheduler in parallel.

\section{Motivation and Attacker Model}
\label{sec:motivation_problem_statement}

\begin{table*}[!tbp]
\caption{Comparison with existing confidential computing mechanisms on specialized accelerator device and their security capabilities.\\
\footnotesize
\centering
\setlength{\tabcolsep}{4pt}
\begin{tabular}{l|l|l|l|l|l}
\req: trusted driver & \partreq: partially trusted driver & \notreq: untrusted driver & \supported: supported & \notsupported: not supported & ?: unknown\\
C-VM: confidential VM & SM: security monitor & HRoT: hardware root-of-trust & SB: Secure boot & RA: remote attestation & LA: Local attestation\\
STA: Single task attestation & TA: task attestation & SC: Security controller & PT: PyTorch & TF: TensorFlow &  
\end{tabular}
}
\begin{adjustbox}{width=2.1\columnwidth, center}
\begin{tabular}{@{}lclcclllllcl@{}}
\toprule
                                            & \multicolumn{5}{c}{CC capability and trust assumption}                                                                                                                                                                                                                                                                                                                                         & \multicolumn{2}{c}{Device}                               & \multicolumn{2}{c}{AI/ML programming capability}                                                                                                           & \multicolumn{2}{c}{Required changes}                                                                                                                                   \\ \cmidrule(l){2-6} \cmidrule(l){7-8} \cmidrule(l){7-8} \cmidrule(l){9-10}\cmidrule(l){11-12}
\multirow{-2}{*}{Existing systems}          & Host TCB                                                                      & \multicolumn{1}{c}{\begin{tabular}[c]{@{}c@{}}Isolation \\ granularity\end{tabular}} & \begin{tabular}[c]{@{}c@{}}Spatial\\ sharing\end{tabular} & \begin{tabular}[c]{@{}c@{}}Multi-residence\\ TEE\end{tabular} & \multicolumn{1}{c}{\begin{tabular}[c]{@{}c@{}}Attestation\\ (CPU/host excl.)\end{tabular}} & \multicolumn{1}{c}{Type} & \multicolumn{1}{c}{Interface} & \multicolumn{1}{c}{\begin{tabular}[c]{@{}c@{}}Native programming\\ interface\end{tabular}} & \multicolumn{1}{c}{AI stack}                                  & HW                                                                           & \multicolumn{1}{c}{SW}                                                                  \\ \midrule
Graviton~\cite{volos2018graviton}           & Intel SGX + \notreq                                                           & \begin{tabular}[c]{@{}l@{}}GPU\\ contexts\end{tabular}                               & \supported                                                & \notsupported                                                 & HRoT,RA,STA                                                                                 & GPU                      & PCIe                          & CUDA                                                                                       & ?                                                             & SC                                                                           & Runtime, drivers, CUDA                                                                   \\
\rowcolor[HTML]{E2E2E2} 
HIX~\cite{jang2019HIX}                      & Intel SGX + \req                                                              & Enclaves                                                                             & \supported                                                & \notsupported                                                 & LA                                                                                          & GPU                      & PCIe                          & CUDA                                                                                       & ?                                                             & \begin{tabular}[c]{@{}c@{}}SGX instruction,\\ MMU, PCIe\end{tabular}         & \begin{tabular}[c]{@{}l@{}}GPU enclave, inter-enclave\\ communication, CUDA\end{tabular} \\
GraphcoreIPU~\cite{vaswani2023confidential} & \notreq                                                                 & Device                                                                               & \notsupported                                             & \supported                                                    & HRoT,SB,RA                                                                                  & IPU                      & PCIe                          & Proprietary                                                                                & TF                                                            & CCU                                                                          & \begin{tabular}[c]{@{}l@{}}XLA, poplar compiler,\\ runtime\end{tabular}                 \\
\rowcolor[HTML]{E2E2E2} 
NvidiaCC (H100)~\cite{h100}                 & C-VM + \req                       & VM                                                                                   & \supported                                                & \notsupported                                                 & HRoT,SB,RA                                                                                  & GPU                      & PCIe                          & CUDA                                                                                       &   TF/PT                                                            & \begin{tabular}[c]{@{}c@{}}Security\\ processor\end{tabular}                 & CUDA, C-VM                                                                             \\
Apple PCC~\cite{pcc}                       & Enclave CPU (?)                                                                             & Node                                                                                 & ?                                                         & ?                                                             & HRoT,SB,RA                                                                                  & NPU                       & Internal                     & Swift                                                                                    & \begin{tabular}[c]{@{}l@{}}Proprietary\\ support\end{tabular} & \begin{tabular}[c]{@{}c@{}}Custom Apple\\ Silicon\end{tabular}               & SepOS, SW stack                                                                         \\
\rowcolor[HTML]{E2E2E2} 
sNPU~\cite{fen2024trusted}                  & \begin{tabular}[c]{@{}c@{}}Penglai Enclave \\ + \partreq{} + SM\end{tabular}       & Worlds                                                                               & \supported                                                & \notsupported                                                 & HRoT,SB                                                                                     & NPU                      & Internal                      & Proprietary                                                                                & ?                                                           &         SoC-NoC, SC                                                                    &    SMMU, SM                                                                                    \\
StrongBox~\cite{deng2022strongbox}          & \begin{tabular}[c]{@{}c@{}}TrustZone\\ + \partreq{} + SM\end{tabular}           & Worlds                                                                               & \notsupported                                             & \notsupported                                                 & SB                                                                                          & GPU                      & Internal                      & OpenCL                                                                                     & ?                                                           & -                                                                            & \begin{tabular}[c]{@{}l@{}}Runtime, driver, \\ MMIO, TASK protector\end{tabular}        \\
\rowcolor[HTML]{E2E2E2} 
Honeycomb~\cite{mai2023honeycomb}           & \begin{tabular}[c]{@{}c@{}}AMD-SEV + \notreq \\ + Validator + SM\end{tabular} & \begin{tabular}[c]{@{}l@{}}SVSM \\ + SM\end{tabular}                                 & \supported                                                & \notsupported                                                 & SB,RA(of SM)                                                                                & GPU                      & PCIe                          & HIP                                                                                        & ?                                                             & -                                                                            & \begin{tabular}[c]{@{}l@{}}Validator, SVSM, \\ SM, runtime\end{tabular}             \\
CAGE~\cite{wang2024cage}                    & \begin{tabular}[c]{@{}c@{}}ARM CCA \\ + \partreq{} + SM\end{tabular}            & Realm                                                                                & \supported                                                & \notsupported                                                 & RA                                                                                          & GPU                      & PCIe                          & OpenCL                                                                                     & ?                                                             & -                                                                            & API, monitors, ShadowTask                    \\
\rowcolor[HTML]{E2E2E2} 
ACAI~\cite{sridhara2023acai}                & \begin{tabular}[c]{@{}c@{}}ARM CCA \\ + \req{} + PCIe port\end{tabular}         & Realm                                                                                & \notsupported                                             & \notsupported                                                 & SB,RA                                                                                       &  \begin{tabular}[l]{@{}l@{}}GPU + \\ FPGA\end{tabular}                      & PCIe                          & CUDA                                                                                       & ?                                                             & -                                                                            & TF-A, SMMU, RMM                           \\
GR-T~\cite{park2023safe}                    & \begin{tabular}[c]{@{}c@{}}Trustzone \\ + \req{} + cloudVM\end{tabular}         & \begin{tabular}[c]{@{}l@{}}VM \\ + Worlds\end{tabular}                               & \supported                                                & \notsupported                                                 & SB,RA                                                                                       & GPU                      & Internal                      & GlobalPlatform API                                                                         & ?                                                            & -                                                                            & Driver Shim, GPU\_shim                      \\
\rowcolor[HTML]{E2E2E2} 
HETEE~\cite{zhu2020hetee}                   & \req{} + SM                                                                    & Node                                                                                 & \notsupported                                             & \notsupported                                                 & SB,RA                                                                                       & GPU                      & PCIe                          & Proprietary                                                                                & ?                                                            & \begin{tabular}[c]{@{}c@{}}HETEE box, PCIe\\ interconnect, (SC)\end{tabular} & SC, API                                                                                 \\
ShEF~\cite{zhao2021shef}                    & \notreq                                                                       & Device                                                                               & \notsupported                                             & \notsupported                                                 & HRoT,SB,RA                                                                                       & FPGA                     & PCIe                          & \notsupported                                                                                                                                             & \notsupported                                                             & -                                                                            & \begin{tabular}[c]{@{}l@{}}ShEF runtime, \\ Shield\end{tabular}                                                                    \\ \midrule
\textbf{\ascendcc}                                   & \notreq                                                                       & Device                                                                               & \notsupported                     & \supported                                                    & HRoT,SB,RA,TA                                                                               & NPU                      & PCIe                          & CANN                                                                                       & PT/TF                                                         & -                                                                            & \begin{tabular}[c]{@{}l@{}}Driver, runtime, \\ kernels (operators)\end{tabular}         \\ \bottomrule
\end{tabular}
\end{adjustbox}
\label{tab:comparison_prior_art}
\end{table*}
%


\myparagraph{Motivation: Gap in the Prior Art}
Large ML/AI workloads involving sensitive and proprietary data require securing data and computation in the cloud \cite{pcc,azure_confidential_cloud}. 
Notably, the rise of LLMs has necessitated confidential computing settings with three parties: the data provider, the cloud provider, and the model provider.
Neither the model nor the data can be leaked to other parties. 
We call this setting \emph{multi-residence TEE} as the TEE needs to access code and data from different mutually distrusting parties. 
This is a clear deviation from the traditional cloud-TEE setting involving two parties: enclave user and untrusted cloud.

Prior works port the confidential computing paradigm to ML-specific accelerators (e.g., NPU \cite{fen2024trusted}, GPU \cite{h100, jang2019HIX, volos2018graviton}). 
As we show in~\Cref{tab:comparison_prior_art}, most existing proposals rely on CPU-based TEE solutions, such as Intel-SGX \cite{volos2018graviton, jang2019HIX}, AMD-SEV \cite{mai2023honeycomb}, TrustZone \cite{deng2022strongbox, park2023safe} or ARM CCA \cite{wang2024cage,sridhara2023acai} on the host, building on it to extend security guarantees across devices. 
This has the advantage of already having a trusted component, i.e., an enclave on the host, to ensure secure communication with the (trusted) device. 
This enables sending authenticated commands using authenticated encryption, signatures, or memory isolation for integrated devices. 
CPU-based TEE approaches significantly enlarge the TCB, requiring trust not just in the accelerator device but also in the CPU and various respective monitors and drivers. Moreover, it complicates compatibility as the solution relies on the specific CPU on the host, and existing attacks~\cite{lipp2018meltdown,foreshadow-usenix18,moghimi2017cachezoom} may even undermine the CPU TEEs' security guarantees.

To the best of our knowledge, only two previous works remove trust from the host completely and solely use the device as the hardware root of trust (HRoT): SheF \cite{zhao2021shef} and Graphcore IPU \cite{vaswani2023confidential}. 
SheF uses an FPGA as the trusted device, ensuring integrity and confidentiality through authenticated and encrypted bitstreams with device isolation. 
Graphcore relies on a specialized compiler to convert the model into an encrypted and authenticated binary in a clean room environment, which can be sent directly to the device without host intervention. 
It requires significant hardware changes and eliminates interactive AI software stacks. 
Therefore, these systems are impractical for real-world, large-scale models.

Our proposal \textit{solely} relies on the NPU as the root of trust, does not require a CPU TEE, works with current AI frameworks, requires no hardware changes, extends to other task-based AI accelerators such as TPUs, and is optimized to run modern real-world LLMs.


\begin{figure}
    \centering
    \includegraphics[width=1\linewidth]{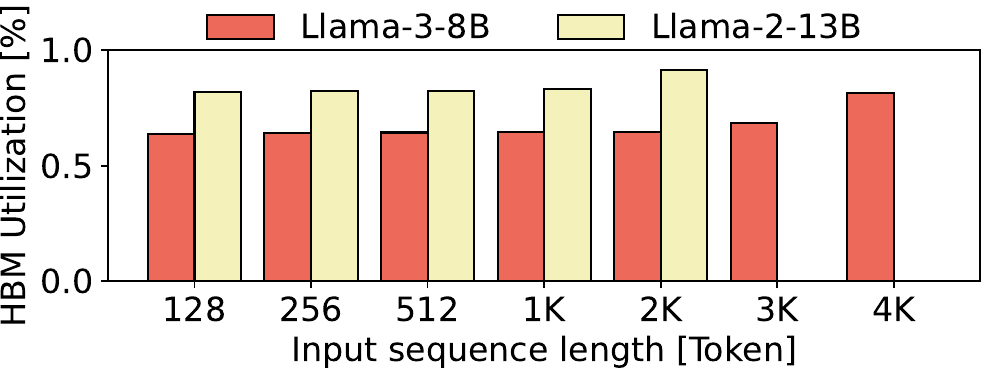}
    \caption{Memory footprint of LLama-3-8B and Llama-2-13B in Ascend 910A NPU with 32GB HBM.}
    \label{fig:memory-footprint}
\end{figure}

\myparagraph{A case against spacial sharing}
The increasing scale of LLMs influences resource-sharing and utilization strategies.
Earlier ML-specific TEEs focused on spatial sharing to boost device utilization. They rely on complex techniques (multiple page tables, monitors, dedicated hardware support) to facilitate resource and performance isolation.
We observe this trend of spatial sharing (also called multi-tenancy) in
commercial confidential computing solutions such as Nvidia-CC on H100 and B100 (using MIG~\cite{mig}),
as well as several academic proposals~\cite{fen2024trusted,wang2024cage,deng2022strongbox, sridhara2023acai,volos2018graviton}. 
Multi-tenancy is often a choice for workloads with low memory utilization, as seen in existing academic proposals such as older CNNs (ResNet, VGG, AlexNet, and MobileNet) or isolated operations such as matrix multiplication or SVD. 
However, such workloads do not represent modern AI workloads like generative AI.
This is evident in~\Cref{fig:memory-footprint}, where we evaluate the memory utilization of state-of-the-art transformers-based LLMs such as Llama2 and Llama3, a Huawei Ascend 910A NPU with 32GB HBM, and we over 90\% HBM utilization.
Moreover, a single NPU has insufficient memory to load a 70B parameter model or execute a 13B parameter model over a 2K input sequence length.
Internal data structures, such as the KV cache, grow quadratic in relation to the input sequence length.
Therefore, increasing sequence length poses a memory capacity challenge. For example, NPU runs out of HMB in Llama-2-13B with an input sequence length of more than 2K.
Secondly, the generative AI applications such as chat-bot~\cite{chatgpt}, code generation~\cite{visualstudioGitHubCopilot}, search~\cite{bingMicrosoftCopilot,perplexity} are latency-sensitive.
Lower compute resources due to spatial sharing between multiple tenants result in a higher latency response. 
Given such a large memory footprint of LLMs and latency \textit{sensitivity}, we conclude that multi-tenancy is \textit{irrelevant}.
Therefore, we explicitly aim for a single-tenant solution.



\myparagraph{Settings} 
The AI workloads involve three parties. 
The model provider brings his IP model, while the data provider uses her secret data to run inference workloads. 
The model provider can also use the data to train or fine-tune the model.
The hardware, i.e., host system, NPUs, network infrastructure, OS, hypervisor, and AI software stack, are deployed by the cloud provider, where all the computations occur.
The cloud provider can also offer the models for the data provider to run an inference service.



\myparagraph{Trust assumption and attacker model}
In a typical trusted execution scenario, the software stack and the cloud provider are untrusted. In addition, our setting involves two types of TEE users: the model and the data provider. The model and data provider are mutually distrusting, and neither trusts the cloud provider:
%
\begin{mylist} 
    \item \textbf{Cloud/infrastructure provider:} The cloud service provider (CSP) is responsible for provisioning and maintaining all hardware and software resources for operation. The CSP controls all the nodes (CPU, NPUs), manages all the infrastructure such as network interfaces, switches, etc., and maintains all the software such as OS, hypervisor, device drivers, firmware, AI/ML software stack such as PyTorch, or TensorFlow. We assume all the hardware and software the cloud provider provides are \emph{untrusted} except the specific NPUs where the AI model execution occurs.
    
    \item \textbf{Model provider:} The model provider develops and trains the model and keeps the model's composition and parameter secret. The model provider may train an open-source model with a proprietary data set. In that case, the parameters of the model are secret. The model provider could also be multiple parties: foundation model owner and fine-tuner. The fine-tuner trains the foundation model with a specific data set to suit application scenarios like video generation or chat. The CSP could also be the model provider in the machine learning as a service (MLaaS) scenario.
    
    \item \textbf{Data owner:} Typically, the data owner is the client who wants to use a specific model and cloud infrastructure (CPU, memory, NPU) for training or inference. This is a scenario where the data owner can bring her model. Typically, we assume that the data and the model provider are two separate, mutually distrusting parties.
\end{mylist}

We only assume the specific NPUs where the AI/ML workloads are deployed are trusted. 
The NPUs have an on-chip hardware security module (HSM) that acts as the hardware root of trust.
Lastly, we assume that denial of service (DoS) and side-channel attacks are outside the scope of this paper.

\section{Security Challenges and Requirements}
\label{sec:motivation_problem_statement:propeties}

\begin{figure}[t]
\begin{lstlisting}[language=iPython]
def mm_npu_kernel(m, n):
    M1 = torch.rand(m,n).npu() #copy M1 to npu
    M2 = torch.rand(m,n).npu() #copy M2 to npu
    M3 = torch.mm(M1,M2).cpu() #copy result M3 to cpu
\end{lstlisting}
\caption{An example code of matrix multiplication on Ascend NPU.}
\label{fig:mm_example_code}
\end{figure}

\begin{figure}[t]
    \centering
    \includegraphics[trim=0cm 13.2cm 17.3cm 0cm, clip,width=0.85\linewidth]{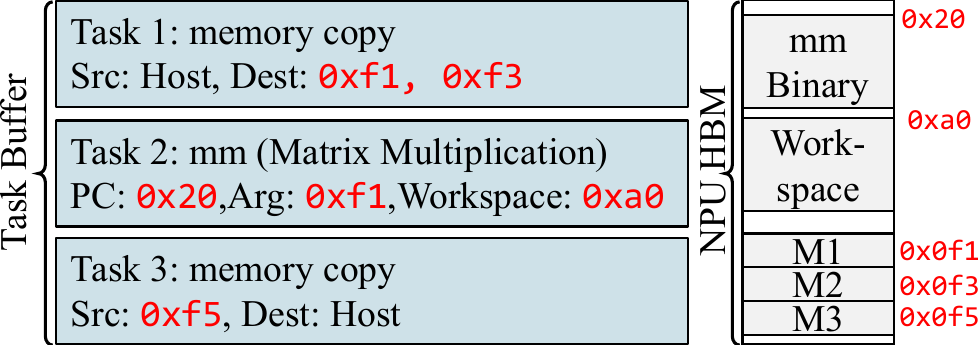}
    \caption{An example matrix multiplication task and memory layout on NPU, corresponding to the code snippet in~\Cref{fig:mm_example_code}.}
    \label{fig:mm_example}
\end{figure}

We use matrix multiplication as a running example, as depicted in~\Cref{fig:mm_example_code}. 
The NPU runtime copies the NPU-optimized binary for matrix multiplication kernel (\texttt{torch.mm}) and \texttt{M1} and \texttt{M2} onto the NPU's memory.
After the kernel executes, the NPU copies the \texttt{M3} from the NPU memory to the CPU's main memory.
The three tasks corresponding to this example are depicted in~\Cref{fig:mm_example}, namely, a memory copy of the tensors (\texttt{M1}, \texttt{M2}) from the host to the NPU, the matrix multiplication and memory copy of the result tensor (\texttt{M3}) from the NPU to the host memory. The host-side NPU runtime reserves memory spaces on the NPU HBM for the binary of the matrix multiplication operator, tensors: \texttt{M1}, \texttt{M2}, \texttt{M3},  and the workspace (operator's working space, e.g., heap and stack). 
Based on this execution model, we observe several security challenges assuming an attacker-controlled host and cloud provider.
Based on these observations, we develop a set of security requirements that the NPU must provide to ensure the security of the model, data, and execution.%

\begin{observation-box-new}{Security Challenge 1}
    The untrusted host runs the privileged software, such as the OS, hypervisor, and device driver, along with the AI software stack, such as PyTorch, that handles the data and model. Therefore, at any point in a traditional AI/ML scenario, the untrusted host has full access to the data and model.%
\label{ob:1}
\end{observation-box-new}
\spacesave
\begin{requirement}
    End-to-end authenticated encryption (such as AES-GCM) is necessary for all input and output tensors, model parameters, and model binaries to ensure the attacker-controlled host cannot observe or manipulate data and models. The host only handles authenticated encrypted (AES-GCM) data, models, and results at any point.%
\label{req:1} 
\end{requirement}

\begin{observation-box-new}{Security Challenge 2}
    The NPU requires the model and data to be decrypted before execution. The attacker-controlled host can extract the data and model when the NPU decrypts the data before execution. Therefore, it is critical to ensure that once the data and model are decrypted inside the NPU, the attacker cannot extract or manipulate the model and data on the NPU. Similarly, end-to-end authenticated encryption of the execution result is not enough, as the host can access the plain text results from the NPU memory before the encryption of the results takes place.%
\label{ob:2}
\end{observation-box-new}
\spacesave
\begin{requirement}
Atomic execution invariant: Before the data and model are decrypted for the model to execute, the host must lose access to the data and model, as the data and model need to be decrypted before execution. This can be achieved by removing the DMA mapping of the NPU memory region where the data and model reside from the NPU side. %
\label{req:2}
\end{requirement}

\begin{observation-box-new}{Security Challenge 3}
The host defines the memory mapping of the model's inputs and outputs. This includes the pointers to the input data and the model and the output result from the model. A malicious host can declare the output pointer to be the same as the input data or the model, compromising the model's confidentiality. %
    \label{ob:3}
\end{observation-box-new}
\spacesave
\begin{requirement}
Memory invariant: For every memory copy from NPU to host, we must ensure that the NPU rejects any memory copy where the plain text input data and model, intermediate results, or output are located.%
\label{req:3}
\end{requirement}

\begin{observation-box-new}{Security Challenge 4}
Even without direct access to the data or the model, the untrusted host can send malicious commands to the NPU, such as copying part of the model parameters to the results sent to the data provider, compromising the model's confidentiality.%
\label{ob:4}
\end{observation-box-new}
\spacesave
\begin{requirement}
The lack of integrity of the model execution compromises the model and data security.%
\label{req:4}    
\end{requirement}

\begin{observation-box-new}{Security Challenge 5}
The security primitives for confidential computing are only valid and secure if there is a mechanism to attest the NPU. Without a systematic way to check the integrity of the NPU firmware (which includes the binaries for the control CPU, task scheduler, NPU memory manager, etc.), we cannot assert the trustworthiness of the NPU's confidential computing capability.%
\label{ob:5}  
\end{observation-box-new}
\spacesave
\begin{requirement}
We need a measured boot-equivalent primitive for the NPU, where the NPU only accepts manufacturer-certified firmware and does not allow the attacker-controlled host to flush its firmware or change the runtime configuration.%
    \label{req:5}    
\end{requirement}

\begin{observation-box-new}{Security Challenge 6}
The NPU and the corresponding software stack provide several debugging methods for correctness and performance, such as inspecting the NPU memory, acquiring a snapshot of a memory region, watching execution time, etc. Such mechanisms allow the attacker to extract models and data. 
\label{ob:6}  
\end{observation-box-new}
\spacesave
\begin{requirement}
To ensure data and model confidentiality, all debugging-related operations must be restricted.%
    \label{req:6}    
\end{requirement}

\section{Basic Building Blocks for Confidential Computing on the Ascend NPU}
\label{sec:buildingBlocks}

Here, we provide the basic design building blocks to enable confidential computing on Ascend NPU that are associated to the requirements described in \Cref{sec:motivation_problem_statement:propeties}.

\subsection{Model and Data Encryption}
\label{sec:buildingBlock:crypto_ops}

 In~\Cref{req:1} (\Cref{sec:motivation_problem_statement}), we describe that it is essential that the AI model and the corresponding data are encrypted by the shared keys between the NPU and model and data providers, respectively. As we assume that the model and data providers are mutually distrusting, they cannot access each other's shared key. 
 
 \myparagraph{Set up} 
 The model and data providers initiate an authenticated key exchange (DH) with the NPU to start a session.
 The NPU derives ephemeral keys from its root key, along with the firmware measurement (derived in the measured boot), with the root key stored in the hardware RoT (refer to~\Cref{sec:buildingblocks:firmware_integrity}).
 This ensures that the model and data provider interact with a legitimate NPU device with the correct firmware image signed by the hardware vendor.

\subsubsection{AI CPU-based custom operator} 

We use NPU AI-CPU cores' ARM AES intrinsic and SIMD (NEON) instructions to accelerate AES-GCM without requiring additional hardware. 
AES-GCM is implemented as an AI CPU operator and can be triggered before (for data and model decryption) and after (for result encryption) the model execution. 
The operators (AI CPU or AI Core) are typically part of the model layer operator; therefore, this eliminates any need for specialized hardware for encryption.
This operator is part of the NPU firmware and is verified by the measured boot during NPU initialization.
All the cryptographic operations are \textit{in-place} and do not need additional memory.

\begin{figure}[!tbp]
    \centering
    \includegraphics[trim=0cm 13.5cm 16.5cm 0cm, clip,width=\linewidth]{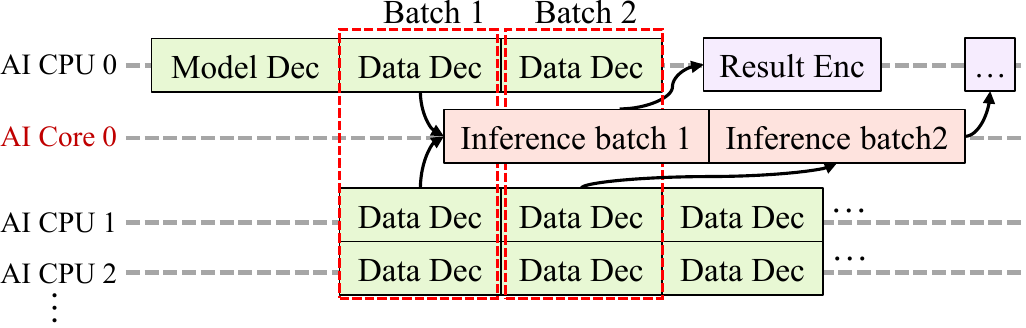}
    \caption{Parallel cryptographic operation on model and data to hide the latency introduced by the AES-GCM operator running on the AI-CPU cores. The AI core executes the AI-related operations, such as the layer computation during an inference pass.}
    \label{fig:model_crypto_parallel}
\end{figure}

The parallelized AES-GCM operator decrypts multiple batches of input data on multiple AI CPU cores to hide the decryption and verification latency.
\Cref{fig:model_crypto_parallel} shows parallel AES-GCM operation on the model, data, and results running on the AI CPUs, while the AI Cores execute the actual AI operations related to the model layers during an inference pass.
Typically, for an LLM, the computation is bound by the inference latency (few milliseconds) compared to the AES-GCM operations for the data (typically in the order of ~100 $\mu$ seconds).
Therefore, the latency is only visible for the first inference, as the model and the first data batch must be decrypted.
The decrypted model will already be in the NPU memory for subsequent inference passes.

\subsubsection{Executing AI CPU operator with model}

\myparagraph{Preparing the model} It starts with the model provider preparing it in a trusted clean room environment. Typically, the compiled model file (also known as the frozen model for the inference) contains the layer information, parameters, and operator binaries. 
The model provider encrypts the weights (a list of \texttt{Tensor}) and the individual operator binaries with the secret key shared between the model provider and NPU beforehand. 
Each binary contains a header, symbol table, and compiled instructions for the NPU AI core.
The model also contains a list of binary sizes for the NPU runtime to use when parsing the model file. 
We modify the sizes to account for the encrypted binary size (added padding) and the 16 bytes added as the message authentication code (MAC).

The model file contains the layer information corresponding to the tasks linked to operator binaries, such as a layer named \texttt{te\_relu\_1\_1}, which denotes the first ReLU activation function.
Typically, a task name corresponds to an actual operation. However, this is not necessary for the model to function correctly.
Therefore, the model provider also replaces the task names with randomized strings to prevent the underlying operator from being exposed.

The encrypted model file is then transferred to the untrusted host, where the AI software stack (e.g., PyTorch) is executed. 
The data provider provides the encrypted data to the untrusted host, encrypted with the secret key shared between the data provider and NPU. 
The host copies the encrypted model and encrypted data to the NPU. 
If the model and data provider are the same party, the model and data are encrypted with the same key. 
After this, the host calls the \texttt{execute()} API to start the inference passes.

\myparagraph{Model execution on the NPU}
The \texttt{executeModel} API call from the NPU runtime indicates the NPU task scheduler to start executing the AI model tasks. 
The NPU task scheduler ensures that the model and data are decrypted right after the \texttt{executeModel} API is called. 
After one inference pass i.e., a forward pass throughout all the layers in the model, the model outputs a vector known as the logits. 
A normalization operation (such as \texttt{softmax}) on the logits produces the probability values of inference classes.
The AES-GCM operator encrypts the logits before they are copied back to the host.

\subsection{Enforcing Memory Lock Invariants}
\label{sec:buildingBlocks:memorylock}

Running the data and model decryption (refer to~\Cref{sec:buildingBlock:crypto_ops}) right after the \texttt{executeModel} API call from the host is insecure. 
It will give the host full access to the decrypted model and data.
Therefore, the host's access must be revoked from these NPU memory regions.
This brings us to \Cref{req:2} and \Cref{req:3}, which are related to critical memory invariants to ensure that the data, model, and execution are isolated from that attacker-controlled host.
We design a memory access control primitive leveraging the SMMU (similar to ARM's SMMU~\cite{arm2013smmu_v3}) on the NPU.
All the PCIe transfers to and from the NPU memory go through the SMMU on the NPU.
Therefore, we enforce access control on the SMMU by reprogramming the NPU control CPU with exclusive access to the NPU-SMMU.
However, there are two challenges that we need to solve to ensure secure access control.

\begin{mylist}
    \item We must ensure that the sequence of events: loading the model to NPU memory $\rightarrow$ locking the NPU memory where the model is loaded $\rightarrow$ decrypting the model, to be atomic, i.e., the host software stack cannot interrupt the NPU.
    \item Once the model and data are decrypted, corresponding virtual memory spaces cannot be unlocked without either re-encryption or resetting the memory content.
\end{mylist}

To solve the first challenge, we ensure that the AES-GCM AI CPU operator for decrypting the model and data can only be scheduled after the NPU memory manager successfully unmaps the model, data, and workspace memory.
Unmapping the memory is achieved using the \texttt{dma\_unmap\_pages} API call, removing all the memory mapping from the SMMU.
By doing so, the NPU memory manager ensures that any memory access from the host is blocked.
Any interrupt from the host at this point will prevent the NPU memory manager from sending an acknowledgment signal to the NPU task manager.
Under normal circumstances, after receiving the acknowledgment signal, the NPU task manager schedules the AES-GCM operator to decrypt and verify the model weights, binaries, and data.

We address the second challenge by introducing a memory exclusivity invariant in the NPU memory manager.
Inside the NPU memory manager, a data structure tracks all the allocated memory locations based on their DMA direction (\texttt{DMA\_BIDIRECTIONAL}, \texttt{DMA\_TO\_DEVICE}, or \texttt{DMA\_FROM\_DEVICE}).
This way, we ensure that the host cannot copy over any data transferred from host to device.
The only memory content that the host is allowed to copy out is the output from the model.
After the model's execution, the NPU task scheduler schedules an AES-GCM encryption operator on the outgoing memory location.
Upon completing the AES-GCM operation, the task scheduler signals the NPU memory manager to remap the memory using \texttt{dma\_remap(location, size)}.
The remap API adds an entry to the SMMU so the host can access and copy the encrypted result.

\begin{figure*}[ht]
    \centering
    \includegraphics[trim=0cm 10.1cm 11.2cm 0cm, clip,width=0.65\linewidth]{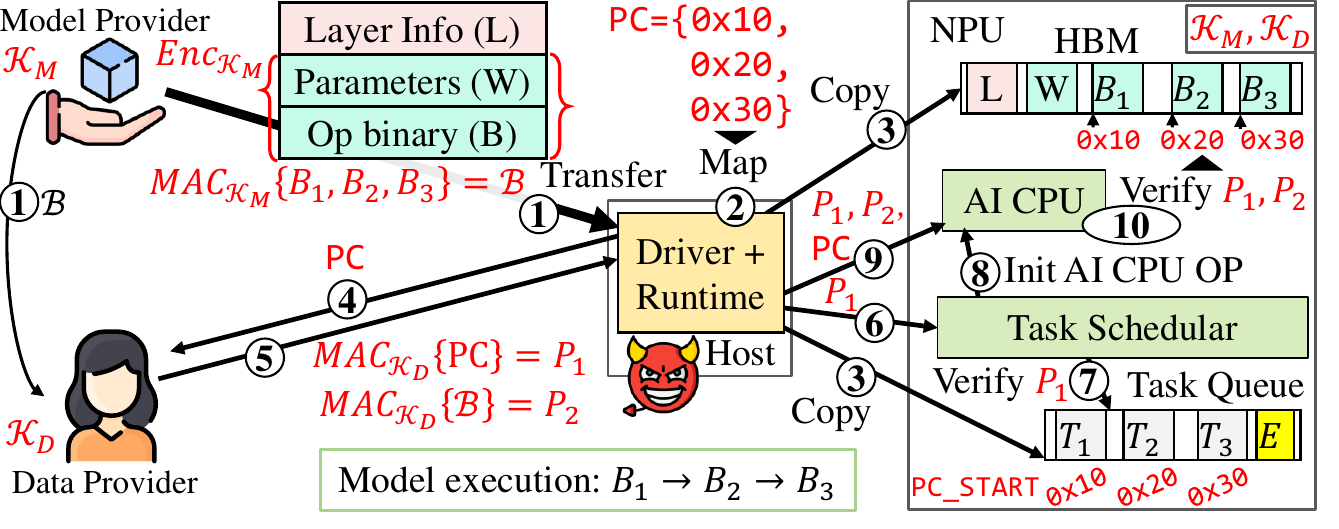}
    \caption{Protocol to ensure integrity and confidentiality of the AI model and integrity of model execution.}
    \label{fig:model_integrityv2}
\end{figure*}

\subsection{Model and Task Attestation}
\label{sec:buildingBlocks:integrity}

\Cref{req:4} implies that the integrity of the model execution is critical to protect the integrity and confidentiality of both model and data.
The model and task attestation mechanism ensures its correct execution. 
 The untrusted NPU driver sends the model file (containing layer information, model parameters, and operator binaries) over DMA, and writes the task to the task queue. 
 The task contains a pointer to the location of the binary (\texttt{PC\_START} attribute) of the specific operator (e.g., matrix multiplication) and the relevant data so that the NPU can execute the code with either the AI CPU or the AI Core.
 As the NPU driver runs on the attacker-controlled host, the host can always push arbitrary tasks and remove or reorder the tasks to compromise the integrity of the AI model.
 The attacker can also modify tasks (\texttt{PC\_START}) to point to a different binary and thus execute a different operator than intended, compromising the integrity of the model execution. To prevent such an attack, we design a model verification method shown in~\Cref{fig:model_integrityv2}. We assume two keys $\mathcal{K}_M$, and $\mathcal{K}_D$ shared respectively between model provider \& NPU, and data provider \&  NPU. 
 We continue with the matrix multiplication example depicted in~\Cref{fig:mm_example_code}, and~\Cref{fig:mm_example} to describe our mechanism.
 Here, the model consists of three layers that execute three operators (memory copy from host to device, matrix multiplication, and memory copy from device to host). The sequence of the corresponding operators' binaries is $B_1\rightarrow B_2\rightarrow B_3$. \Cref{fig:model_integrityv2} depicts the flow of our model and task attestation mechanism. The steps are the following:
 \begin{mybullet} 
     \item[\one] An AI model consists of layer information (L), model parameter (W), and operator binaries (B = $B_1,B_2,B_3$) to execute a matrix multiplication of two tensors of shape $\{2,2\}$. 
     The model provider encrypts and creates message authentication codes (MAC) of W and B. Alongside this, the model provider also generates the MAC of the sequence of binaries corresponding to the layers, such as $\mathcal{M} \leftarrow Enc_{\mathcal{K}_M}(Model)$ and $\mathcal{B}\leftarrow MAC_{\mathcal{K}_M}(B_1, B_2, B_3)$. 
     Typically, the layer information contains the name of the layer (that the model provider can masquerade) and the location of the encrypted binary relative to a fixed starting point in the model file so that the NPU runtime can create corresponding layer tasks later. 
     The model provider sends $\mathcal{M}$ and $\mathcal{B}$ to the untrusted host (on the CSP) and $\mathcal{B}$ to the data provider.
     
     \item[\two] The host deploys the model to the NPU. 
     The runtime determines the starting addresses of binaries ($B_1,B_2,B_3$) known as the \texttt{PC} to map the binaries on the memory based on the available space on the HBM. In our example, \texttt{PC=\{0x10,0x20,0x30\}}.
     
     \item[\three] The driver copies the model file over to the NPU memory over DMA. 
     As~\Cref{fig:model_integrityv2} shows, after the memory copy, the HBM contains the layer information, the model parameters, and the binaries. 
     At this point, W and $\{B_1,B_2,B_3\}$ are encrypted with $\mathcal{K}_M$. 
     At the same time, the NPU runtime creates a sequence of tasks: $T_1, T_2, T_3$ from the layer information and uses \texttt{PC} to populate the \texttt{PC\_START} attribute.
     
     \item[\four] The untrusted host sends the \texttt{PC} to the data provider.
     
     \item[\five] In response, the data provider generates     
       $P_1\leftarrow MAC_{\mathcal{K}_D}(\texttt{PC}), P_2\leftarrow MAC_{\mathcal{K}_D}(\mathcal{B})$, 
       where $\texttt{PC}\leftarrow\{\texttt{0x10,0x20,0x30}\}, \mathcal{B}\leftarrow MAC_{\mathcal{K}_M}(B_1, B_2, B_3)$, and sends them to the host.  
     
     \item[\six] 
     The driver sends the \texttt{executeModel} instruction to the task scheduler by writing a specific execution task ($E$ in~\Cref{fig:model_integrityv2}) on the task queue. 
     $P_1$ is sent together with  \texttt{executeModel}
     This triggers the task scheduler to remove all the memory mapping from the NPU's SMMU. 
     Therefore, the host can no longer read or write from and to either the HBM or the task queue.
      
      \item[\seven] The task scheduler collects all the \texttt{PC\_START} attributes from the task queue. The task scheduler calculates $P'_1\leftarrow MAC_{\mathcal{K}_D} (\texttt{PC\_START})$ and checks if $P_1= P'_1$. Otherwise, the task scheduler aborts.
     
     \item[\eight] If the verification in Step \seven is successful, the task scheduler invokes an AI CPU operator to inspect the integrity and sequence of binaries.
     
     \item[\nine] The host send $P_1, P_2$, and \texttt{PC} to the AI CPU operator. 
     
     \item[\ten] After receiving $P_1,P_2$, and \texttt{PC}, the AI CPU operator recomputes $P_1$ (with $\mathcal{K}_D$) from \texttt{PC} and checks if it matches the $P_1$ received from the untrusted host.
     Using \texttt{PC}, the AI CPU decrypts the binaries one by one, each time checking that the binaries match their associated MAC. Once the binaries are all decrypted, the AI CPU computes $\mathcal{B}'\leftarrow MAC_{\mathcal{K}_M}(B_1, B_2, B_3)$ and $P_2'=MAC_{\mathcal{K}_D}(\mathcal{B}')$. It checks that $P_2'=P_2$. If all of these steps are successful, the AI
CPU operator then decrypts W in place on the HBM.
     Note that removing all memory mappings in the NPU SMMU in step \six prohibits the untrusted host from reading the decrypted W and B on the HBM or writing new content on the HBM. 
 \end{mybullet}

\subsection{Firmware and Runtime Integrity}
\label{sec:buildingblocks:firmware_integrity}

The previously stated mechanisms are implemented in the NPU firmware, which is part of \ascendcc{} software TCB. 
Therefore, the trustworthiness of \ascendcc{} depends on the integrity and authenticity of the firmware.
The NPU has a hardware root of trust, where the NPU vendor embeds a cryptographic key during the manufacturing process (e-fuse keys).
The root key cannot be extracted from the NPU.
It serves as the NPU's unforgeable identity, preventing the attacker from impersonating and emulating a legitimate NPU.
Subsequent keys for key exchanges are derived from this root key and signed with the NPU vendor's key.
The control CPU boots and sets up the NPU, including the PCI drivers, task scheduler, NPU memory manager, AI CPU, and AI Cores.
During the boot, the control CPU verifies whether the firmware image was signed with the manufacturer's root key.
This prevents the attacker from flashing an unsigned firmware image to the NPU.
We assume the cloud service provider has a public key infrastructure to ensure that the model and data provider can execute authenticated Diffie-Hellman key exchange with the NPU to derive shared secrets, encrypt and authenticate the AI model and data, and verify the legitimacy of the NPU.

The NPU control CPU intercepts all the command messages from the host-side runtime. 
\ascendcc{} blocks all debugging commands (e.g., memory inspection, profiling operators) from the control CPU to ensure the attacker has no additional communication channel with the NPU.

\section{\ascendcc}
\label{sec:ascendcc}

\begin{figure*}[t]
    \centering
    \begin{minipage}{.7\textwidth}
        \centering
        \includegraphics[trim=0cm 11.2cm 7.6cm 0cm, clip,width=1\linewidth]{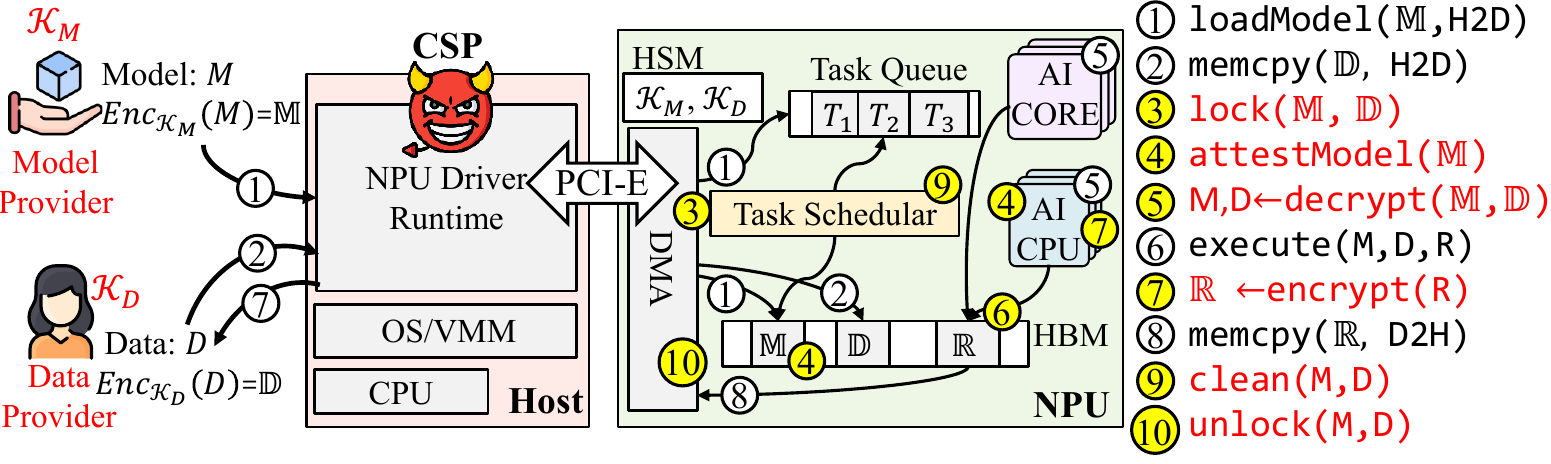}
    \end{minipage}%
    \begin{minipage}{.4\textwidth}
        \includegraphics[trim=-2cm 9.4cm 23cm 0cm, clip,width=0.8\linewidth]{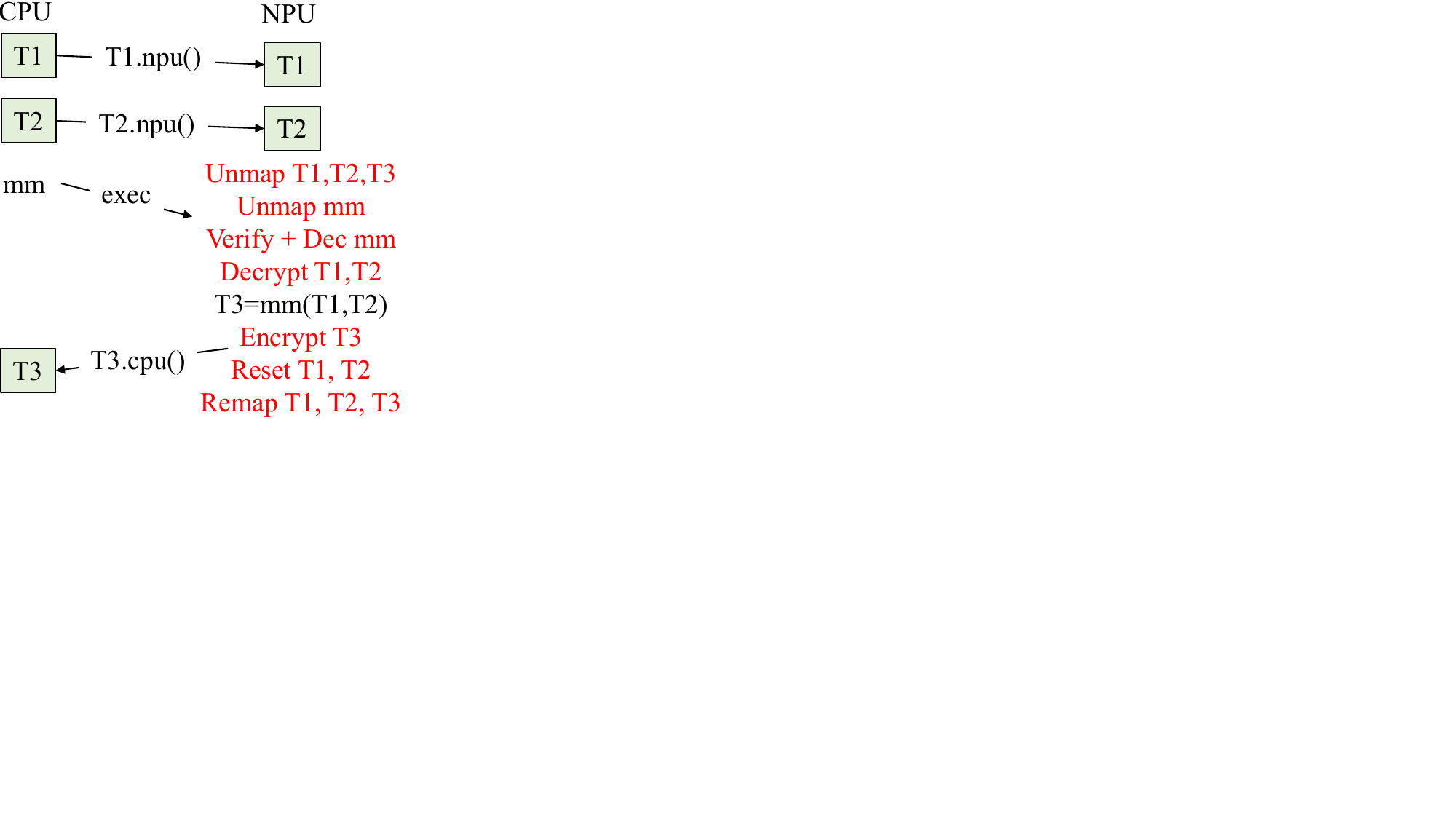}
    \end{minipage}%
    \caption{\ascendcc{} end-to-end system with internal confidential computing mechanisms and the corresponding PyTorch interfaces.}
    \label{fig:systemdesign}
\end{figure*}

Based on the fundamental design building blocks discussed in~\cref{sec:buildingBlocks}, we briefly describe the \ascendcc{} end-to-end system. \ascendcc{} works seamlessly with \texttt{torch\_npu}~\cite{githubGitHubAscendpytorch} which is a NPU-specific adapter for PyTorch. 

\myparagraph{Initial setup} The model and data providers execute an authenticated Diffie-Hellman key exchange (DHKE) with the NPU over the untrusted host and obtain $\mathcal{K}_M$ and $\mathcal{K}_D$, respectively. 
The NPU stores these two keys on the HSM on-chip.
The data provider prepares encrypted data with the shared secret $\mathcal{K}_D$.
Assume the data provider interacts with the remote LLM application through a browser.
A browser extension loaded with $\mathcal{K}_D$ can tokenize the data provider's input ($D$), encrypt it ($\mathbb{D} \leftarrow Enc_{\mathcal{K}_D}(D)$), and send the encrypted tokens to the cloud provider.
Similarly, the model provider encrypts the model binaries and parameters and generates the encrypted model $\mathbb{M}\leftarrow Enc_{\mathcal{K}_M}(M)$. The model provider also generates the signed sequence of model binaries for model attestation as described in~\Cref{sec:buildingBlocks:integrity}.

\myparagraph{\ascendcc{}: End-to-end system} \Cref{fig:systemdesign} shows end-to-end \ascendcc{} systems along with the sequence of operations (step \circled{1} to step  \circled{10}). 
The figure also shows where the individual steps take place.
The highlighted steps are security-sensitive steps that \ascendcc{} adds to the NPU firmware to enable confidential computing.
A brief summary of the steps follows:  \circled{1} the host calls \texttt{loadModel} to send the encrypted model $\mathbb{M}$ to the NPU. 
 \circled{2} Then \texttt{memcpy} transfers the encrypted data $\mathbb{D}$.
At this point, the host calls \texttt{executeModel} API that instructs the NPU to start executing the AI model with the data.
In \ascendcc{}, \texttt{executeModel} triggers additional steps to ensure that the model and data are inaccessible from the attacker-controlled host after the decryption and that the model tasks are not manipulated.
\circledy{3} \texttt{lock}$(\mathbb{M},\mathbb{D})$: The NPU task scheduler intercepts the \texttt{executeModel} from the host and instructs the NPU memory manager to remove the model, workspace, and data region, located on the NPU HBM, from the SMMU mapping. 
Therefore, the model, data, and workspace on the NPU are no longer accessible from the host.
This mechanism is described in~\Cref{sec:buildingBlocks:memorylock}.
\circledy{4} \texttt{attestModel} $(\mathbb{M})$: The NPU task scheduler verifies the tasks and model binaries. This is a multi-step protocol that involves a dedicated AI CPU operator. The model and task attestation mechanism is described in~\Cref{sec:buildingBlocks:integrity}.
\circledy{5} Once the memory regions are locked, the task scheduler invokes an AI CPU operator to decrypt the model and data (refer to~\Cref{sec:buildingBlock:crypto_ops}) in place. 
\circled{6} The AI model is executed on the AI cores and AI CPUs based on the operators in the model layers. The output of the model is $R$. The memory region(s) where $R$ resides is not DMA-mapped to the host.
\circledy{7} The NPU task scheduler invokes the AI CPU crypto operator (\Cref{sec:buildingBlock:crypto_ops}) to encrypt the model output with the data provider's key: $\mathbb{R} \leftarrow Enc_{\mathcal{K}_D}(R)$)
\circled{8} The \texttt{memcpy} API from the host copies encrypted output $\mathbb{R}$ from the NPU memory to the host memory.
\circledy{9} If the host runs another inference pass, the task scheduler invokes a specialized AI CPU operator to zero out the previous input data. If the host triggers an end of the session, the task scheduler cleans up both the model and data.
\circledy{10} The task scheduler invokes the NPU memory manager to remap the memory (either $D$ or both $D,M$) before the start of the next inference pass or the end of the session.

\myparagraph{Programming interface}
\ascendcc{} modifications in the NPU firmware and driver such as the model and task attestations, cryptographic operators, and memory invariant enforcement are completely transparent to higher-level programmings, such as the AI software stack (PyTorch).
Therefore, from the developer's perspective, the existing inference (and even training) workflow remains unchanged. 
\ascendcc{} simply acts as a drop-in-replacement for the AI developers.
There are minimal changes within the PyTorch adapter for Ascend NPU (i.e., \texttt{torch\_npu}) to support the collection of model attestations. 
\Cref{fig:systemdesign} shows the matrix multiplication example (from~\Cref{fig:mm_example_code}) with all the \ascendcc{} hardening.

\begin{figure}[t]
    \centering
    \includegraphics[trim=0cm 11.5cm 20.7cm 0cm, clip,width=0.8\linewidth]{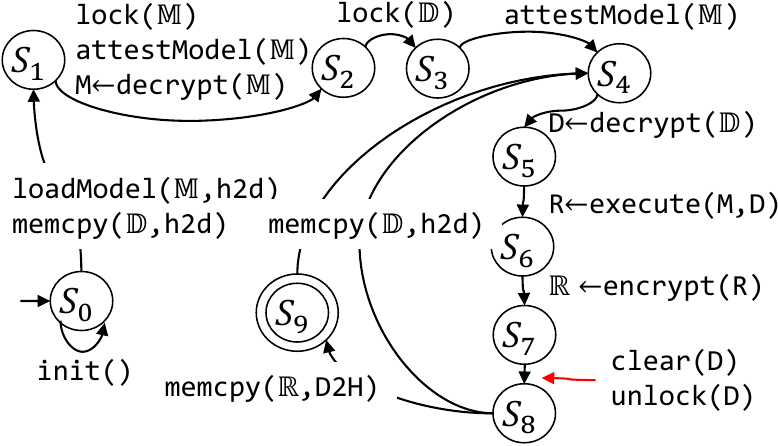}
    \caption{\ascendcc{} memory  and execution lifecycle.}
    \label{fig:lifecycle}
\end{figure}

\myparagraph{\ascendcc{} lifecycle}
We summarize the memory lifecycle of \ascendcc in ~\Cref{fig:lifecycle} by providing the memory states over subsequent rounds of inference passes.
Typically, the model is loaded once, followed by multiple inference passes (such as a conversation with an LLM chatbot).
Therefore, the memory region associated with the model (parameters, binary, and model operator workspace) must be locked and decrypted once. 
Then, it remains locked for the duration of the runtime until the model is unloaded or an interrupt occurs.
The data, on the other hand, arrives in a streaming fashion.
The data region is locked, decrypted, and fed to the model execution.
Once the output is generated, it is encrypted with the data owner's key, and the corresponding memory region is unlocked for device-to-host DMA transfers.
At the same time, the input regions are reset (by overwriting them to zero using a dedicated AI CPU operator), and the corresponding memory regions are unlocked. 
Consequently, the host can transfer the next batch of encrypted input data to the NPU.

\section{Security Analysis}
\label{sec:security_analysis}

In this section, we provide an informal security analysis of \ascendcc{} and show how it ensures the security of AI models and data from the untrusted host and cloud provider.

\myparagraph{Attacks from a malicious host and cloud provider}
The host runs the operating system/hypervisor, the NPU driver, and the AI software stack and has full access to the NPU.
The host needs to have initial access to allocate and copy over memory for the data, model, and operator binaries.
Once the host calls the \texttt{execute()} API, the NPU task scheduler unmaps the NPU memory.
Therefore, the host cannot access the encrypted model and data anymore.
The host can interrupt the NPU between the memory locking and the model and data decryption to disrupt the DMA region locking.
However, before the NPU task scheduler schedules the AI CPU operator to decrypt the model and data, a successful confirmation from the NPU's memory manager is required to ensure that the memory is not accessible from the host.
Similarly, the host can interrupt the NPU between the result encryption and memory unlocking to prevent the result encryption from occurring.
However, the NPU task scheduler only triggers the NPU memory manager to unlock the DMA memory after successfully executing the AI CPU operation.
The task scheduler obtains this confirmation from its completion queue (CQ), indicating if an operator has completed successfully or failed.
This atomic property prevents the host from accessing the plaintext model and data on the NPU memory.
The host cannot issue a malicious DMA to modify the model and data as they contain the MAC from the model and data provider.
The host cannot manipulate the tasks after the \texttt{executeModel} is called as the memory is locked.
The NPU control CPU has disabled all the debugging and performance monitoring interfaces to prevent the host from having any additional communication channel with the NPU.
As all data entering and leaving the NPU is encrypted and authenticated, a malicious cloud provider with physical access cannot compromise the device's security.
Note that denial of service (DoS) is always possible and out-of-scope of \ascendcc{}.
The malicious cloud provider can flash an NPU card with an older or compromised firmware.
However, as described in \Cref{sec:buildingblocks:firmware_integrity}, the NPU collects the firmware measurement during the measured boot.
This signed information will be passed on to the model and data provider during the key establishment.
The model and data provider can detect if the NPU runs an older or compromised firmware version.
If the attacker tries to boot an emulated NPU, it will fail, as it does not have the private NPU key.

\myparagraph{Attacks from a malicious model provider}
The motive of a malicious model provider is to steal data from the data provider. 
A malicious model provider can manipulate the AI operator code to execute unintended operations. For example, it can copy part of the data to the model output and retrieve it later.
We ensure all output data is encrypted with the data provider's key.
Therefore, only the data provider can decrypt the output from the model.

\myparagraph{Attacks from a malicious data provider}
A malicious data provider tries to steal model parameters or infer operator code, known as a model stealing (MS) attack.
The data provider also tries to infer the training data by sending many queries to the model, known as a membership inference attack (MIA).
Typically, in any confidential computing solution, MS and MIA are orthogonal problems, so they are considered out-of-scope.
The model provider can deploy additional measures in the operators to add noise to the input or reject inference after a given number of queries.
The model provider ensures that these security mechanisms are in place via the model and task attestation (refer to~\Cref{sec:buildingBlocks:integrity})
This makes both MIA and MS prohibitively expensive for the data provider.



\section{Implementation and Evaluation}
\label{sec:results}

\subsection{\ascendcc{} Implementation}

We implement all our confidential computing primitives into Ascend 910A NPU's software stack, which involves the driver, firmware, runtime, and Ascend PyTorch adapter written in C++. All encryption and decryption are executed with AES-GCM-128 based on the AArch64cryptolib~\cite{cryptolib} library that uses ARM's hardware cryptographic intrinsic for fast performance. For LLMs, we use both the Ascend native execution (acl runtime) and the PyTorch adapter to execute the models. For GPT-Neo, we use the ImageNet dataset, and for transformers, we use the squad\_v2 dataset~\cite{DBLP:journals/corr/abs-1806-03822}.

\myparagraph{Model provider}
We emulate the model provider by implementing it as a TCP server that serves authenticated-encrypted compiled models with a shared secret negotiated previously with the NPU. 
The MAC of the parameters and operator instructions are appended after each encrypted blob, increasing the size of each binary by 16 bytes. 

\myparagraph{Host runtime} \label{host_runtime}
We modify the NPU runtime (CANN)~\cite{CANN} to implement the model and task attestation described in~\Cref{sec:buildingBlocks:integrity}.
During the \texttt{loadModel} API call, we extract the tasks and associated \texttt{PC\_START} attribute pointing to the operator binary's memory location.
The modified runtime expects an additional signed binary sequence from the model file, establishing the model's ground truth and task attributes.
The modified runtime communicates with the data provider over the TCP socket to retrieve the signed \texttt{PC\_START} sequence.

\myparagraph{Task scheduler}
The modified task scheduler (TS) firmware enforces the memory invariant and attestation.
The TS firmware records all the tasks submitted since the last execution completion. 
Upon reception of the \texttt{executeModel} command, the TS firmware computes the signature of the \texttt{PC\_START} sequence using $\mathcal{K}_D$, as described in \Cref{host_runtime}. It then verifies that the signature matches the one generated by the data provider and sent with the \texttt{executeModel} command. If all the steps above are successful, the TS firmware relays the \texttt{executeModel} command to the TS hardware.

\myparagraph{AI CPU operators}
We use Huawei Ascend NPU AI-CPU operator development environment~\cite{ai-cpu-dev} that is part of the Ascend AI development software stack known as CANN~\cite{CANN} to implement the custom AI CPU operator.
The custom operators are responsible for AES-GCM decryption and encryption (refer to~\Cref{sec:buildingBlock:crypto_ops}) for the model, data, and results and carrying out model and task attestation(refer to~\Cref{sec:buildingBlocks:integrity}). 
The AI CPU operators are executed whenever they are called from a compiled model graph.
They could also be called native C++ kernels or be from PyTorch.
All the AI CPU operators are implemented as internal operators, i.e., the compiled binary of the operator remains inside the trusted NPU firmware. Therefore, a user cannot modify the operator code.

\myparagraph{Memory lock}
%
We use the \texttt{dma\_map\_page} and \texttt{dma\_unmap\_page} functions to isolate the device from the host before starting the execution (\Cref{sec:buildingBlocks:memorylock}).
The host only accesses mapped HBM regions on the device via DMA using shared virtual memory (SVM).
Removing the mapping of the corresponding DMA addresses prevents unauthorized host access. 
The NPU task scheduler sends a synchronized message (implemented over shared memory, non-accessible to the host) to the memory manager driver running on the NPU control CPU, which then uses \texttt{dma\_unmap\_page} to unmap the entire mapped region. 
Only after the unmapping does the task scheduler start with the model and task attestation (\Cref{sec:buildingBlocks:integrity}). 
In case the unmapping fails for any reason, the device aborts the execution. 
The task scheduler stores the input address and length to clear and remap the memory region for the next inference. 
During the remapping, the encryption operator informs the task scheduler when it has encrypted the output, informing the SVM driver that respective regions can be remapped so that the output can be read (and the next input can be loaded). 

\subsection{LLM Evaluations}

We chose two different types of LLM workloads based on their parameter size: GPT-Neo, a small model with 125 million parameters, and large models, Llama2 (7 and 13 billion) and Llama3 (8 billion).
Typically, in large models, the inference time is higher (seconds). Therefore, the relative overhead of \ascendcc{} is small.
However, the inference time is within a second in smaller models like GPT-Neo.
Such a fast inference speed magnifies the overhead the different \ascendcc{'s} modules introduce.
First, we start with microbenchmarks, where we evaluate different parts of \ascendcc{}, and after that, we present end-to-end LLM overheads.

\begin{figure}[!tbp]
    \centering
    \includegraphics[width=1\linewidth]{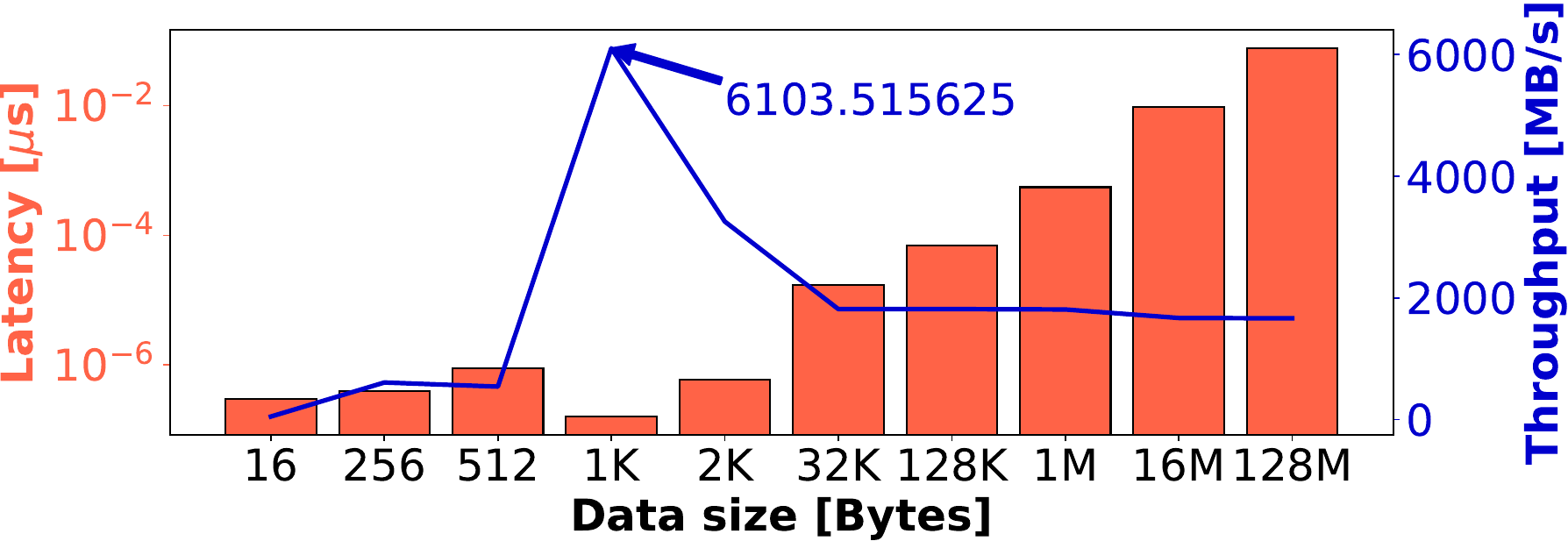}
    \caption{AES-GCM-128 operator latency on the Ascend 910A AI-CPU cores with auto-tiling for concurrency. 
    }
    \label{fig:aes-gcm-performance}
\end{figure}

\subsubsection{Micro benchmarks}
The AES-GCM AI-CPU operator has an effect on both the setup and inference time. 
\Cref{fig:aes-gcm-performance} shows the AES-GCM-128 performance on all four AI-CPU cores on Ascend 910A NPU.
We use the NPU's automatic scheduling strategy to maximize parallelization.
The four AI-CPU cores have a shared 1KB L2 cache.
Therefore, we achieve a maximum throughput of 6.1 GB/s when the data size is exactly 1KB and is spread evenly across all 4 cores (i.e., 256B for each core).
For data sizes larger than 1K, the cores experience cache contention and converge to single-core performance, around 1.6 GB/s. 
Therefore, in our AI-CPU operator, we always execute on 1KB of maximum chunk size (with interleaved DMA transfer) to ensure maximum throughput rate.

\begin{figure}[!tbp]
    \centering
    \includegraphics[width=1\linewidth]{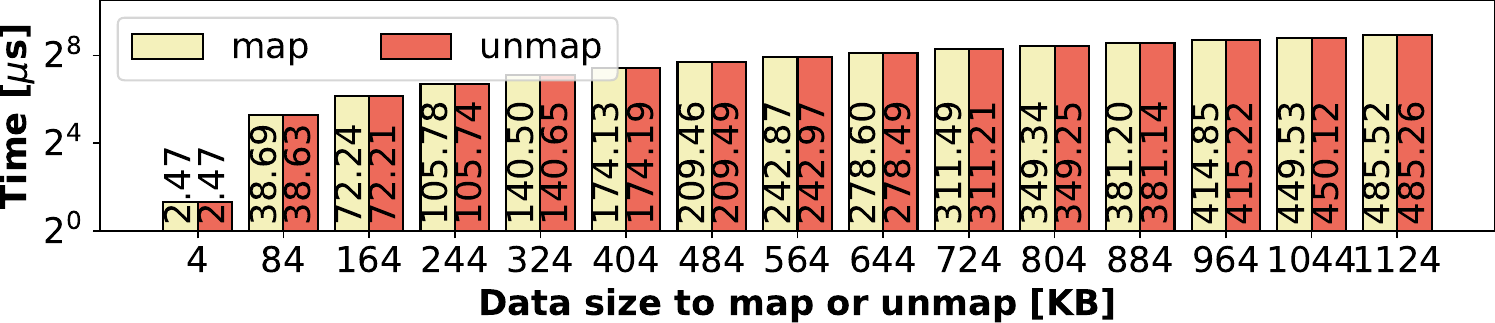}
    \caption{Time (overhead) to map and unmap the Ascend 910A VAs from host VAs. 
    }
    \label{fig:map-unmap-all}
\end{figure}

\Cref{fig:map-unmap-all} shows the latency to map (\texttt{dma\_map\_pages}) or unmap (\texttt{dma\_unmap\_pages}) call from the NPU memory manager to either remap or unmap DMA memory region for the host.
The map and unmap operations are at the granularity of pages of size 4K, which takes 2.47 $\mu$s.
However, most unmap operations are done during the model setup to unmap the model and the model workspace from the host.
In the subsequent inference passes, \ascendcc{} only needs to unmap the input data before decryption, map the encrypted result, and remap the input region after resetting the previous input.

\subsubsection{LLM Inference and Model Setup Evaluation}

We evaluate the \ascendcc{} setup time and runtime overhead.
The setup time is a one-time cost that occurs when the model is loaded on the NPU and prepared for inference passes.
Runtime overhead denotes the added latency for token generation during inference time.
Typically, the user only notices the runtime overhead while interacting with the LLM application, such as the chatbot or the code completion assistant.

\begin{figure}[!tbp]
    \centering
    \ifPPI
\else
        \includegraphics[width=0.8\linewidth]{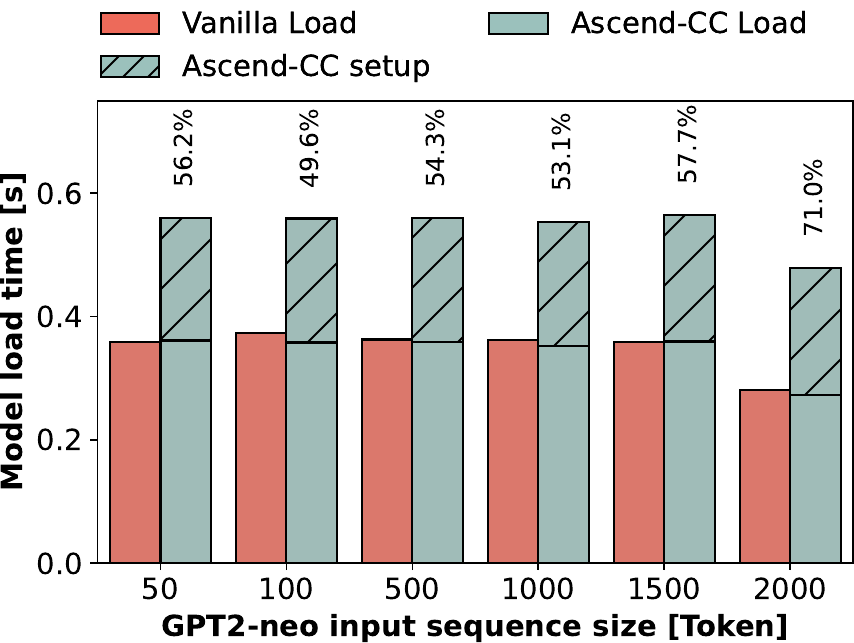}
    \fi
    \caption{GPT-Neo-125 Load time overhead. The solid color box indicates the time to execute \texttt{loadModel}, i.e., the DMA transfer time of the model file. The hatched boxes show additional setup time for \ascendcc{}.
    }
    \label{fig:gpt-load-overhead}
\end{figure}

\begin{figure}[!tbp]
    \centering
    \includegraphics[width=1\linewidth]{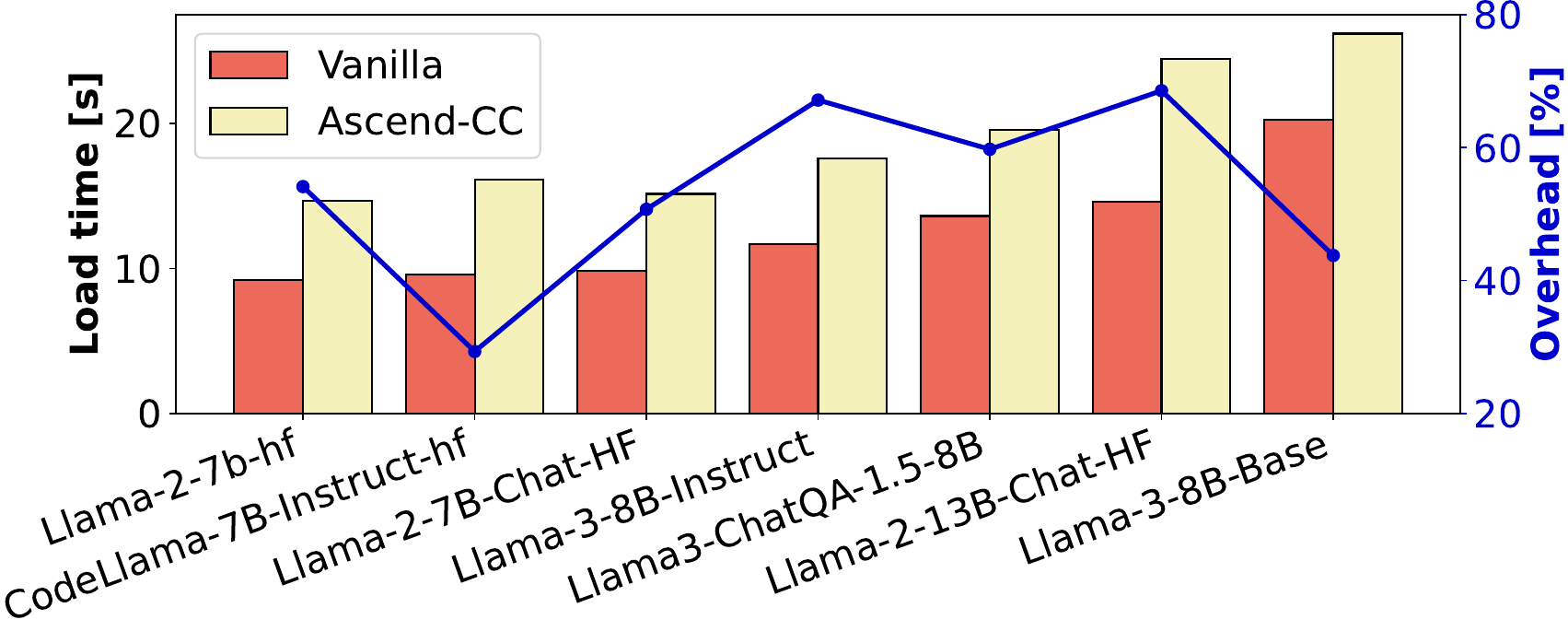}
    \caption{Llama models load time overheads (\%) in \ascendcc{}. 
    }
    \label{fig:llama-load-overhead}
\end{figure}

\myparagraph{Setup overhead} We provide set-up costs for two scenarios: a small model such as GPT-Neo with 125 million parameters and large models such as Llama-2 and Llama 3 with 7, 8, and 13 billion parameters.
As expected, the smaller model has significantly lower loading times (e.g., 0.25 seconds in GPT-Neo vs 26 seconds in Llama-3 8B) than the larger models.
However, due to the shorter loading time, the pay-per-inference has a noticeable effect on the small model. At the same time, we did not observe any measurable difference in the Llama-2 and Llama-3 variants.
In GPT-Neo, the overhead of setup time is between 50 to 71\%, as seen in~\Cref{fig:gpt-load-overhead}.
A similar trend can also be seen in the Llama2 and Llama3 evaluations depicted in~\Cref{fig:llama-load-overhead}, where we evaluate the setup time of 7 LLMs. 
Typically, the setup time reflects the size of the AI model, which is proportional to the model parameter count.
We also observe that Llama-3-8B-Base has the highest load time in vanilla as the model is encoded with \texttt{bfloat16} data type.
Ascend 910A NPU does not have native support \texttt{bfloat16}. Hence, it is required to convert the datatype from \texttt{bfloat16} to standard \texttt{float16}.
This datatype conversion introduces additional latency in loading.
The rest of the models are encoded in native \texttt{float16} and do not require additional typecasting before loading the model.
Therefore, the loading times are proportional to the model's parameter count.

\begin{figure}[!tbp]
    \centering
    \includegraphics[width=1\linewidth]{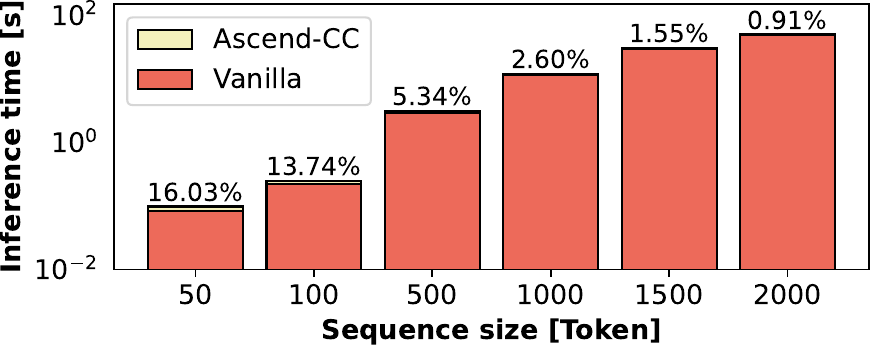}
    \caption{Inference time overhead (\% on the bars) of GPT-Neo-125M with different input sequence sizes. 
    }
    \label{fig:gpt-inference-overhead}
\end{figure}

\begin{figure*}[!tbp]
    \centering
    \subfloat{{\includegraphics[width=7cm]{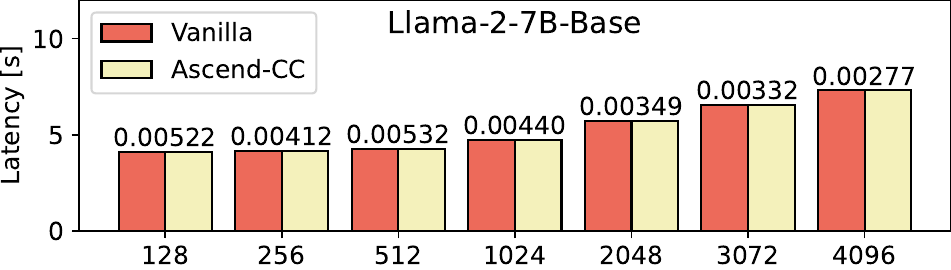} }}%
    \qquad
    \subfloat{{\includegraphics[width=7cm]{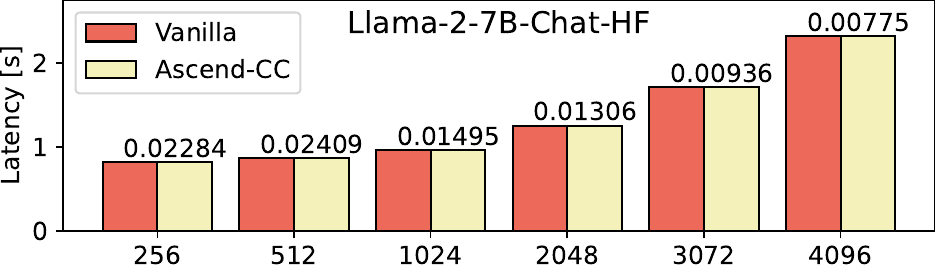} }}%
    \qquad
    \subfloat{{\includegraphics[width=7cm]{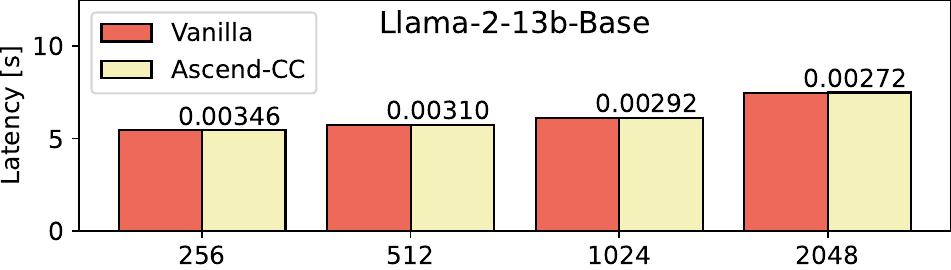} }}%
    \qquad
    \subfloat{{\includegraphics[width=7cm]{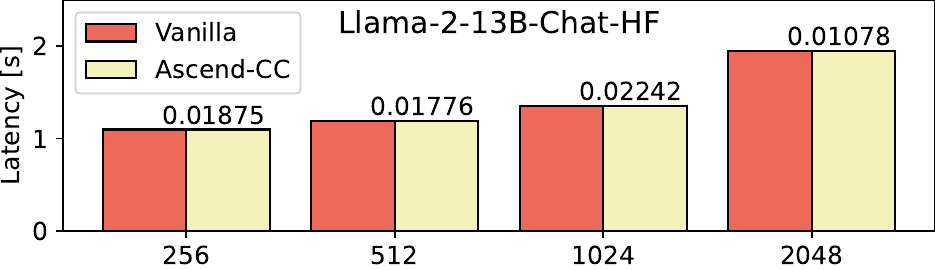} }}%
    \qquad
    \subfloat{{\includegraphics[width=7cm]{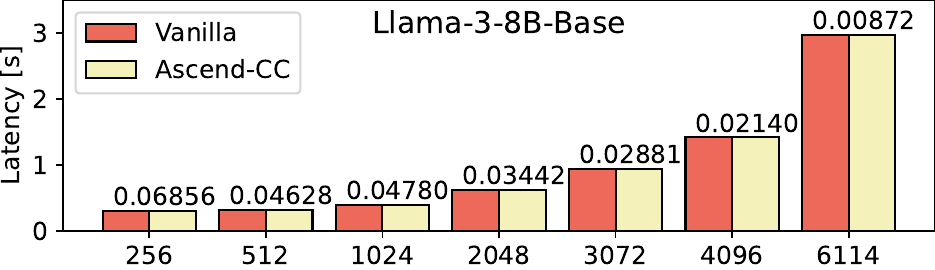} }}%
    \qquad
    \subfloat{{\includegraphics[width=7cm]{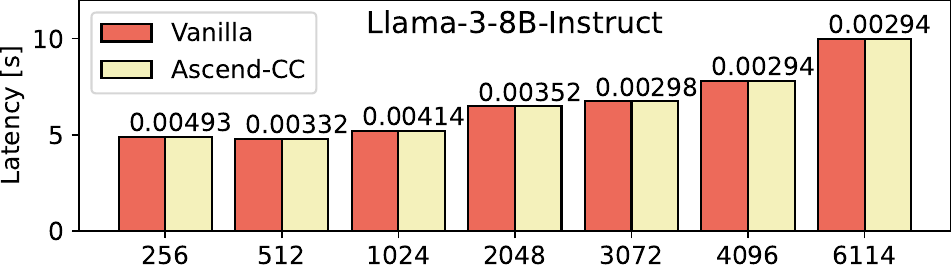} }}%
    \qquad
    \subfloat{{\includegraphics[width=7cm]{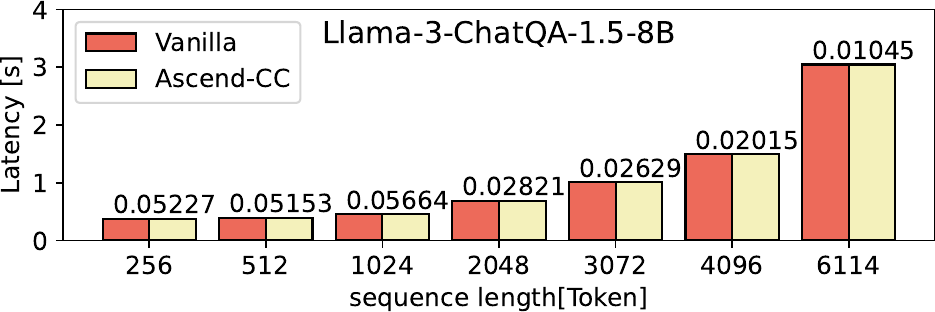} }}%
     \qquad
    \subfloat{{\includegraphics[width=7cm]{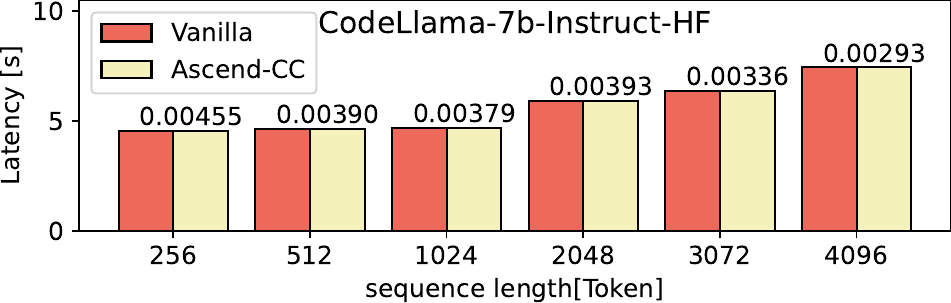} }}%
    \caption{Llama-2 inference overhead in different input sequence sizes. We show a single inference cost for both vanilla and \ascendcc{}. The number on the bar indicates the overhead of \ascendcc{} in percentage (\%). 
    }%
    \label{fig:llama2-inference}%
\end{figure*}

\myparagraph{Inference overhead}
\Cref{fig:gpt-inference-overhead}, and~\Cref{fig:llama2-inference} show the runtime overhead of \ascendcc{} with GPT-Neo 125M. In the case of GPT-Neo, the overhead is significantly higher compared to Llama models, as the model is much smaller (125M vs. 7/8/13 B). 
Therefore, the inference latency is very small.
In very input lengths such as 50 and 100 tokens, we experience 16.03\% and 13.74\% overhead, respectively.
However, a small input length (typically less than 512) is impractical in real-world LLM applications.
Larger sequence lengths result in higher inference latency.
For a 2K context size, \ascendcc{} only introduces 0.91\% overhead.

\ascendcc{'s} overhead reduction in LLMs with a higher parameter count is apparent in models such as Llama-2, Llama-3, and CodeLlama, along with their variants (chat, instruct, and Q\&A). 
These results are depicted in~\Cref{fig:llama2-inference} for different context sizes.
Note that we only show up to a 2K context size for the 13B variant of Llama-2 as larger input sequence sizes in these two models caused out-of-memory errors. 
We observed a reduction of \ascendcc{} overhead with larger input sequence sizes.
This is expected as the inference latency with increasing sequence sizes is dominated by the model computation rather than the cryptographic operations.
We generally observe less than 0.1\% overhead in all Llama variants across all input sequence sizes.
Note that in all of these experiments, we use a batch size of one. With larger batch sizes, the overhead increases by a small fraction.



\section{Related Work}
\label{sec:related_work}

Several works aim to secure sensitive data from the model providers by running sensitive computation inside a CPU TEE (e.g., Intel SGX). While some methods run the entire model on the CPU TEE \cite{lee2019occlumency}, this undermines the advantages of specialized ML task accelerators. Therefore, other approaches try to partially mitigate this shortcoming by only running selected parts of the workload inside the CPU-TEE while utilizing accelerators for more intensive tasks \cite{mo2020darknetz, tramer2018slalom, hashemi2021darknight}. These approaches, though, still have a significant overhead and are vulnerable to privacy-stealing attacks \cite{zhang2023no}.

In parallel (and as detailed in \cref{sec:motivation_problem_statement}), there has been a long line of work aimed at directly extending the CPU-TEE or C-VMs to specific accelerator devices \cite{cronus, ren2023accshield, jang2019HIX, sridhara2023acai, h100}.

Closest to this paper, works like Graphcore \cite{vaswani2023confidential} and SheF \cite{zhao2021shef} move the trust entirely from the host to the device used to run the workload. Similarly, GuardNN\cite{hua2022guardnn} removes the trust from the host by redesigning an FPGA as an entirely new secure accelerator for ML tasks. 
On a larger scale, a few approaches \cite{zhu2020hetee, dhar2022empowering} try to extend the confidential computing paradigm to data-center architectures, allowing many devices to be split across users. 
Complementary to the aforementioned techniques, several works purely focus on improving memory isolation through improved capabilities \cite{woodruff2014cheri, yu2023capstone}, or enhancing I/O isolation mechanisms for confidential computing \cite{feng2024siopmp}, which often directly benefits TEE-devices architectures.

In contrast, some proposals aim to preserve privacy algorithmically either via accelerator-enabled secure multi-party computation (MPC) using secret sharing \cite{tan2021cryptgpu, kumar2020cryptflow} or homomorphic encryption \cite{juvekar2018gazelle, gilad2016cryptonets}. Despite significant progress in these fields, both solutions still incur significant overhead when applied to many practical, real-world workloads \cite{mishra2020delphi}, making them impractical for our setting.

While using a TPM for remote attestation has been the most common approach, there have been a few examples for software-based attestation \cite{seshadri2004swatt, ibrahim2016darpa}, mainly for IoT devices.  
Upcoming PCIe features enable mechanisms to connect CPU-TEEs with DSA-TEEs. 
Specifically, TEE Device Interface Secure Protocol (TDISP) for PCIe-6 enables TEEs on processors to connect to the TEE-enabled PCIe accelerator~\cite{pcie6}. 
Integrity and Data Encryption (IDE) on PCIe-5 encrypts and integrity protects PCIe traffic on processors and devices~\cite{pcie5-IDE}. 
The adoption of these PCIe extensions as TDX-Connect for Intel TDX, SEV-TIO for AMD SEV-SNP, Device Attach (DA) for Arm CCA or IOPMP for RISC-V allows these TEE-enabled devices to benefit from a secure direct access to the TEE memory on processors~\cite{tdx-connect, amd-sev-tio, arm-da, IOPMP}. 

While side channels are out of scope, some works have proposed solutions to secure workloads against such attacks when using a TEE. For instance, Telekine \cite{hunt2020telekine} secures workloads for such TEEs as presented in Graviton \cite{volos2018graviton}.
\section{Conclusion}
\label{sec:conclusion}

We present \ascendcc{}, a system to enable confidential computing for large language models that ensures security against a strong attacker model, including the host CPU.
\ascendcc{} provides memory invariants and cryptographic operators leveraging the heterogeneous architecture of NPUs.
We implement \ascendcc{} on a Huawei Ascend910A NPU and evaluate it on state-of-the-art generative AI workloads.
Our evaluation shows that \ascendcc{} is practical and protects the model and data from the untrusted host.

\bibliographystyle{unsrt}
\bibliography{bibliography} 
\end{document}